\documentclass[aps,superscriptaddress,twocolumn,a4paper,prd]{revtex4}
\usepackage[colorlinks=true, pdfstartview=FitV, linkcolor=blue, citecolor=red, urlcolor=black]{hyperref}
\usepackage{graphicx}
\usepackage[all]{xy}
\usepackage{amsmath}
\usepackage{amssymb}
\usepackage{tensor}
\usepackage{orcidlink}
\newcommand{\be}{\begin{equation}}
\newcommand{\ee}{\end{equation}}
\newcommand{\ben}{\begin{eqnarray}}
\newcommand{\een}{\end{eqnarray}}
\newcommand{\bes}{\begin{subequations}}
\newcommand{\ees}{\end{subequations}}
\def\bal#1\eal{\begin{align}#1\end{align}}

\newcommand{\bfi}{\begin{figure}}
\newcommand{\efi}{\end{figure}}
\newcommand{\bc}{\begin{center}}
\newcommand{\ec}{\end{center}}

\newcommand{\sech}{\mbox{sech}}

\newcommand{\LL}{{\cal L}}

\newcommand{\Ac}{{\cal A}}

\newcommand{\Mc}{{\cal M}}
\newcommand{\Tc}{{\cal T}}

\begin{document}

\title{Generalized scalar field models in the presence of impurities}
\author{D. Bazeia\,\orcidlink{0000-0003-1335-3705}}
     \email[]{bazeia@fisica.ufpb.br}
    \affiliation{Departamento de F\'\i sica, Universidade Federal da Para\'\i ba, 58051-970 Jo\~ao Pessoa, PB, Brazil}
    
        \author{M.A. Marques\,\orcidlink{0000-0001-7022-5502}}
        \email[]{marques@cbiotec.ufpb.br}
    \affiliation{Departamento de Biotecnologia, Universidade Federal da Para\'iba, 58051-900 Jo\~ao Pessoa, PB, Brazil}
    
    \author{R. Menezes\,\orcidlink{0000-0002-9586-4308}}
     \email[]{rmenezes@dcx.ufpb.br}
    \affiliation{Departamento de Ci\^encias Exatas, Universidade Federal da Para\'iba, 58297-000 Rio Tinto, PB, Brazil}

\begin{abstract}
We study generalized scalar field models coupled to impurities in Minkowski spacetime with arbitrary dimensions. The investigation concerns a class of models that depends explicitly on the spacetime coordinates and also, it reveals the presence of a second-order tensor that can have null divergence if a first-order equation and a constraint are satisfied. We obtain the conditions to get compatibility between the equation of motion and the first-order equation, within a framework that is also used in the static case, to show that the introduction of an auxiliary function may allow to describe the energy density of the solution as a divergence. Stability of the solution under rescale of argument, translation in the space and small fluctuations are also fully investigated. We further illustrate the procedure considering the canonical model and also, the $k$-field and Born-Infeld-like models. The results show that stable solutions can be obtained in arbitrary dimensions, and the stability seems to be related to the first-order equation that emerges from imposing null divergence of the aforementioned tensor.
\end{abstract}

\maketitle

\section{Introduction}
Scalar fields are of interest in several branches of Physics. For instance, they can be used in the study of localized structures in Field Theory \cite{manton} and magnetic domain walls in Condensed Matter \cite{fradkin}. Usually, localized structures arise under the action of scalar and other fields. Kinks are the simplest ones and appear in models of a single real scalar field in $(1,1)$ spacetime dimensions. The canonical model which supports these structures is given by a Lagrangian density formed by the difference between kinetic and potential terms \cite{vachaspati}. The equation of motion that governs the kink engenders nonlinearity and is of second order. The kink solutions are static and connect the neighbor minima, degenerated in energy, of the potential. The model is invariant under translations in spacetime, so one can define a conserved energy-momentum tensor \cite{landau}. This allows that we calculate the energy density and follow the so-called BPS procedure \cite{bogo,ps} by finding the conditions that lead to solutions which minimize the energy of the system. This method leads to first-order equations whose solutions (kinks) are stable, compatible with the equations of motion.

The study of localized structures in $(D,1)$ flat spacetime dimensions has a caveat shown by Derrick's theorem \cite{derrick}. By performing a rescale of argument in the scalar field solution, it was shown that no stable localized configuration can be obtained in two or more spatial dimensions governed by canonical scalar field model. However, there are some ways to evade this issue. For instance, one can consider non-canonical models of the $k$-field type, in which $k$-defects \cite{babichev} arise, allowing for the presence of stable solutions in $(D,1)$ dimensions \cite{kglobal}. Also, one may study a model consisted of a complex scalar field coupled to a gauge field via a $U(1)$ local symmetry to get stable vortex solutions in $(2,1)$ flat spacetime dimensions \cite{no}. In the same line, there is the possibility of investigating models with $SU(2)$ symmetry coupling triplets of scalar and gauge fields to obtain magnetic monopole \cite{thooft,polyakov} and dyon \cite{juliazee} solutions in $(3,1)$ flat spacetime dimensions.

Among the many possibilities to circumvent Derrick's theorem, one may consider the presence of the spatial coordinate in the Lagrangian density, i.e., the presence of spatial inhomogeneities in the system usually called \emph{impurities}. This idea was investigated in Refs.~\cite{prl,morris1} with the inclusion of the explicit dependence on the radial coordinate in the potential term of the Lagrangian density, allowing for stable radially symmetric solutions in arbitrary dimensions. The presence of impurities may be useful for more practical investigations, in which the spatial uniformity may not be taken for granted. Over the years, several systems with impurities have been investigated in the literature, including vortices \cite{vimp1,vimp2,vimp3,vimp4,vimp5,vimp6,vimp7}, Bose-Einstein condensates \cite{becimp1,becimp2,becimp3,becimp4} and Fermi liquids \cite{fermi1,fermi2}.

In the specific case of kink-like structures, the study of models in the presence of impurities has been conducted since the decade of 1970; see Refs.~\cite{imp1,imp2,imp3,imp4,imp5,imp6,imp7,imp8,imp9,imp10,imp11,imp12,imp13,imp14,imp15,imp16,imp17,imp18,imp19,imp20}. The explicit presence of the spatial coordinates in the Lagrangian density associated to the aforementioned models breaks translational invariance. Therefore, one cannot obtain a conserved energy-momentum tensor anymore. However, considering that both the impurity and the field are static, one can define an energy density associated to the solutions. This was used in Refs.~\cite{imp10,imp11,imp12,imp13}, in which impurity-doped models of a single real scalar field in $(1,1)$ spacetime dimensions engendering the BPS property, i.e., minimum energy configurations, were introduced. Later, in Ref.~\cite{imp17}, it was shown that, considering the presence of static impurities, it is still possible to obtain BPS solutions in arbitrary dimensions. The study of kink collisions in the presence of impurities has also gained interest recently \cite{scattering1,scattering2,scattering3,scattering4} as it leads to interesting features, such as the appearance of spectral walls, which may induce the bounce of the defect.

The study of impurities with scalar fields is currently done by considering the canonical model with the inclusion of terms that couple the impurity with the derivative of the field and the potential. However, non-canonical (or generalized) models, in which the Lagrangian density is an arbitrary function of the kinetic term and the scalar field, are of current interest and have been widely considered in the literature; see Refs.~\cite{tach1,brax,tach2,tach3,tach4,babichev,trilogia1,adamk,trilogia2,superk,altw,twin1,twinadam1,twinadam2,zhongk,superktrodden,stabkink}. In particular, in Ref.~\cite{tach1} the so-called tachyonic dynamics was introduced, which couples the potential with the dynamical term via a product. Generalized models also give rise to other interesting properties, such as the possibility of obtaining twinlike models \cite{altw,twin1,twinadam1,twinadam2}, which are noncanonical models that support the same solutions and energy densities of the canonical correspondent models. Non-canonical models also allow for a more comprehensive investigation of scalar field models, unveiling more general features, such as the Derrick's argument, which was shown to require the stressless condition \cite{trilogia1}, and the linear stability, which is described by a Sturm-Liouville eigenvalue equation that can be factorized in terms of adjoint operators \cite{trilogia2,stabkink}.

This work is an extension of a Letter \cite{letterimpurity} devoted to the same subject, but there we proposed to study mainly the canonical model. Here, we investigate how to construct generalized scalar-field models coupled to impurities in arbitrary spacetime dimensions via an additive term in the derivative of the field and a specific product with a function of the field itself. We develop a new formalism, based on the null divergence of a second-order tensor similar to the energy-momentum tensor, seeking for the presence of first-order equations compatible with the equations of motion for non-static configurations. We also investigate the conditions under which static solutions are stable against rescale of argument, spatial translations, and small fluctuations, unveiling how to calculate the energy in terms of the integral of a divergence. In Sec.~\ref{sec2}, we develop the general aspects of our procedure, including the conditions under which the above mentioned second-order tensor engenders null divergence even with the explicit presence of spacetime coordinates in the Lagrangian density. This leads us to a surprisingly novel possibility, so we have further considered a specific class of models, showing that a first-order equation can be obtained. Moreover, in order to stress the main result, we investigate the important possibility of considering static field and impurities, emphasizing the stability and energy of the field configurations. In Sec.~\ref{secexamples}, we illustrate the procedure with several distinct systems, in particular, the canonical, $k$-field and Born-Infeld-like models, seeking for analytical results. In Sec.~\ref{secfinal}, we conclude our investigation with some final remarks and perspectives for future works.

\section{Scalar field models with impurities}\label{sec2}

Before considering the study of coordinate-dependent Lagrangian densities, let us first review the action associated to a single scalar field $\phi$ without explicit dependence on the spacetime coordinates in $(D,1)$ flat spacetime dimensions, with metric tensor $\eta_{\mu\nu}=\text{diag}(1,-1,-1,\ldots,-1)$. We can write
\be\label{ac1}
S = \int dt\,d^Dx\,\LL(\phi,\partial_\mu\phi).
\ee
As it is well-known, the above Lagrangian density can be associated with the following generalized momentum 
\be\label{momentum}
\Pi^\mu = \frac{\partial\LL}{\partial(\partial_\mu\phi)}.
\ee
By varying the action in Eq.~\eqref{ac1} with respect to the field, we get the equation of motion
\be\label{eom}
\partial_\mu\Pi^\mu = \LL_\phi,
\ee
with $\LL_\phi = \partial\LL/\partial\phi$. Since the action \eqref{ac1} is invariant under spacetime translations, $x^\mu\to x^\mu + a^\mu$, where $a^\mu$ is constant, Noether's theorem ensures that the energy-momentum tensor $\tensor{T}{^\mu_\nu} = \Pi^\mu\partial_\nu\phi - \delta^\mu_\nu\LL$ is conserved, i.e., $\partial_\mu \tensor{T}{^\mu_\nu}=0$ if the equation of motion \eqref{eom} is satisfied. Evidently, this is a consequence of the absence of explicit dependency on the spacetime coordinates in the Lagrangian density.

Let us now suppose that the action takes into account the presence of impurities via explicit dependence on the spacetime coordinates, $x^\mu$, of the form
\be\label{actionxmu}
S = \int dt\,d^Dx\,\LL(\phi,\partial_\mu\phi,x^\mu).
\ee
In this situation, the vector $\Pi_\mu$ defined in \eqref{momentum} depends on $\phi$, $\partial_\mu\phi$ and $x_\mu$, but the equation of motion \eqref{eom} remains valid. However, since $x^\mu$ is explicitly present in the above expression, the invariance under spacetime translations that gives rise to the energy-momentum tensor is lost, and the energy-momentum tensor is not conserved anymore. Actually, the tensor $\tensor{T}{^\mu_\nu}$ defined below Eq.~\eqref{eom} is no longer a true energy-momentum due to the presence of impurities. For example, the explicit time dependence makes it difficult to define the energy of the system. However, as we will show in Sec.~\ref{staticgeneral}, interesting features arise when the impurity is static, which is the proper scenario to study localized structures. Moreover, in Sec.~\ref{secstaticD1} we will focus on the case of a single spatial dimension, which is special due to requirements concerning stability of the field configuration.

To avoid misinterpretations with respect to the energy-momentum tensor $\tensor{T}{^\mu_\nu}$ associated to the impurity-free model \eqref{ac1}, from now on we will refer to the tensor
\be\label{emt}
\tensor{\Tc}{^\mu_\nu} = \Pi^\mu\partial_\nu\phi - \delta^\mu_\nu\LL
\ee
for the impurity-doped model \eqref{actionxmu}. By using the equation of motion \eqref{eom} for the action \eqref{actionxmu} and taking the divergence of \eqref{emt}, we get $\partial_\mu \tensor{\Tc}{^\mu_\nu} + \LL_{x^\nu} =0$, where $\LL_{x^\nu}$ represents the derivative with respect to the \emph{explicit} dependence on $x^\nu$. Even though $\tensor{\Tc}{^\mu_\nu}$ is not conserved, we can force it to have null divergence with the condition
\be\label{cond}
\LL_{x^\nu}=0.
\ee
Of course, if the coordinates do not appear explicitly in the Lagrangian density, the above equation is automatically satisfied for all the solutions of the equation of motion and $\tensor{\Tc}{^\mu_\nu}$ recovers the conserved energy-momentum tensor $\tensor{T}{^\mu_\nu}$. However, as we shall show below, other possibilities compatible with above condition lead to interesting properties, even if the system does not support a conserved energy-momentum tensor due to the presence of impurities.

Indeed, we have found an interesting class of models that engender explicit dependence on the coordinates via impurities and still allows for a divergenceless $\tensor{\Tc}{^\mu_\nu}$. It is given by the Lagrangian density
\be\label{lagrangian}
\LL = F(\phi,X) + \sigma_\mu G^\mu.
\ee
In this expression, $F$ is, in principle, an arbitrary function of the field $\phi$ and $X$, which will be defined below. Also, $\sigma_\mu=\sigma_\mu(x^\alpha)$ is the impurity vector, which describes the presence of impurities that contain explicit dependence on the spacetime coordinates. In this system, the impurities enter the model independently of the field configurations, so they do not represent an auxiliary field. Also, $G_\mu=G_\mu(\phi)$ is another vector, which only depends on the scalar field. In this work we will suppose that both $\sigma_\mu$ and $G_\mu$ are vectors belonging to the Minkowski spacetime, with the spatial portions described within the Cartesian basis. Also, some component of $G_\mu$ only exists if its corresponding $\sigma_\mu$ also exists. Thus, if one takes $\sigma_0=0$, for example, the existence of $G_0$ makes no sense anymore and one must take $G_0=0$. The dynamical term associated to the scalar field in the presence of impurities is represented by $X$, and here it has the form
\be\label{xdef}
X = \frac12\left(\partial_\mu\phi + \sigma_\mu\right)\left(\partial^\mu\phi + \sigma^\mu\right).
\ee
 In this new situation, the generalized momentum \eqref{momentum} takes the form
\be\label{momentuml}
\Pi^\mu = F_X\left(\partial^\mu\phi + \sigma^\mu\right),
\ee
where $F_X=\partial F/\partial X$. Notice that the above momentum depends explicitly on $\sigma^\mu$, and the equation of motion \eqref{eom} associated to \eqref{lagrangian} now reads
\be\label{eoml}
\partial_\mu\Pi^\mu = F_\phi + \sigma_\mu G^\mu_\phi,
\ee
in which we have used the notation $F_\phi=\partial F/\partial \phi$ and $G^\mu_\phi = dG^\mu/d\phi$.
The components of the tensor \eqref{emt} for the Lagrangian density \eqref{lagrangian} are
\bes\label{emtcomp}
\begin{align}
\tensor{\Tc}{^0_0}&=F_X\left(\dot{\phi} + \sigma_0\right)\dot{\phi} - F - \sigma_\alpha G^\alpha,\\
\tensor{\Tc}{^0_i}&=F_X\left(\dot{\phi} + \sigma^0\right)\partial_i\phi,\\
\tensor{\Tc}{^i_0}&=F_X\left(\partial^i\phi + \sigma^i\right)\dot{\phi},\\
\tensor{\Tc}{^i_j}& = F_X\left(\partial^i\phi + \sigma^i\right)\partial_j\phi - \delta^i_ j\left(F + \sigma_\alpha G^\alpha\right),
\end{align}
\ees
where the dot represents derivative with respect to time. Since the Lagrangian density \eqref{lagrangian} engender explicit dependence on the spacetime coordinates due to the presence of $\sigma_\mu$, it is not invariant under spacetime translations and we cannot get a conserved energy-momentum tensor anymore. However, we can work with the tensor \eqref{emt} to ensure that it has null divergence. The condition required to achieve this is \eqref{cond}, which leads to
\be
\left(\Pi_\mu + G_\mu\right)\partial_\nu\sigma^\mu=0.
\ee
This equation is satisfied by $\Pi_\mu + G_\mu=0$, which can be combined with Eq.~\eqref{momentuml} to give 
\be\label{fogeral1}
F_X\!\left(\partial_\mu\phi + \sigma_\mu\right) + G_\mu =0.
\ee
This is a differential equation of first order that is easier to solve than the equation of motion \eqref{eoml}. However, we must ensure that the above equation is compatible with \eqref{eoml}, so we must also require
\be\label{constraint}
\frac{G^\mu_\phi G_\mu}{F_X} = F_\phi.
\ee
Notice that the above equation involves $\phi$ and $X$. Nevertheless, from the first-order equation \eqref{fogeral1}, one can write
\be\label{XG}
2XF_X^2 = G_\mu G^\mu,
\ee
which is an algebraic equation that relates $X$ and $\phi$, since $F_X$ may depend on $\phi$ and $X$, and $G^\mu$ depends only on $\phi$. When possible, the above equation admits a solution in the form $X=H(\phi)$. This makes the equation \eqref{constraint} become a constraint that dictates how the functions of $\phi$ must appear in the Lagrangian density. The first-order equation \eqref{fogeral1} can be written as
\be\label{fogeral}
\partial_\mu\phi = - \sigma_\mu -\frac{G_\mu}{F_X\big|_{X=H(\phi,G^\mu)}} .
\ee
It is worth to remark that, even though the constraint \eqref{constraint} ensures that the solutions of the first-order equation \eqref{fogeral1} are also solutions of the equation of motion \eqref{eoml}, not all solutions of the equation of motion satisfy the first-order equation. We can only be sure that the solutions of the equation of motion which lead to the null divergence of $\tensor{\Tc}{^\mu_\nu}$ also solve the first-order equation.

\subsection{Static case}\label{staticgeneral}
The formalism introduced in the previous section is valid for time-dependent fields and impurities. However, we can further explore properties of the model if we consider static system. This includes an important class of models, with $\phi=\phi(\vec{x})$ and the Lagrangian density in \eqref{actionxmu} now having the form $\LL=\LL(\phi,\nabla\phi,\vec{x})$. In this situation, the components of the tensor \eqref{emt} are
\bes
\bal
\tensor{\Tc}{^0_0} &=-\LL, \\
\tensor{\Tc}{^0_i} &= \tensor{\Tc}{^i_0}=0,\\
\tensor{\Tc}{^i_j} &= \Pi^i\partial_j\phi-\delta^i_j\LL.
\eal
\ees
Since the impurity is static, we have a well-defined energy density with the component $\tensor{\Tc}{^0_0}$. By integrating it in all space, one obtains the energy, i.e.,
\be\label{energy}
E=\int_{\mathbb{R}^D} d^Dx\, \tensor{\Tc}{^0_0}.
\ee
We can follow the lines of Ref.~\cite{letterimpurity} and use the above expression to study Derrick's scaling argument and investigate the stability under contractions and dilations by taking $\vec{x}\to\vec{y}=\lambda\vec{x}$, which implies $\phi(\vec{x})\to\phi^{(\lambda)}(\vec{x})=\phi(\lambda\vec{x})$, with $\lambda$ being a real parameter. We denote the energy of the rescaled solution as $E^{(\lambda)}$. By imposing that the minimum energy occurs for $\lambda=1$, we require that $\partial E^{(\lambda)}/\partial\lambda|_{\lambda=1}=0$ to get
\bes\label{derrick1}
\be\label{derrick1a}
\int_{\mathbb{R}^D} d^Dx\,\tensor{\Tc}{^i_i}=0,
\ee
where the index $i$ has to be summed over. We also require that $\partial^2 E^{(\lambda)}/\partial\lambda^2|_{\lambda=1}>0$, to obtain
\be\label{derrick1b}
\begin{aligned}
    \int_{\mathbb{R}^D} d^Dx\big(&2D\Pi^i\partial_i\phi-D(D+1)\LL + 2\Pi^i_{ x^j}\,x^j\partial_i\phi\\
   & -\frac{\partial^2\LL}{\partial(\partial_i\phi)\partial(\partial_j\phi)}\,\partial_i\phi\partial_j\phi-\LL_{x^ix^j}x^ix^j\big)>0.
\end{aligned}
\ee
\ees
where the subindices in the coordinates, $x^i$, stand for the derivative with respect to their explicit dependences. To obtain Eqs.~\eqref{derrick1a} and \eqref{derrick1b}, we have used Eq.~\eqref{cond} to ensure the null divergence of $\tensor{\Tc}{^\mu_\nu}$. If the solutions satisfy these conditions, they are stable under contractions and dilations. 

We may also investigate stability under spatial translations, $\vec{x}\to\vec{y}= \vec{x} + \vec{k}$, where $\vec{k}$ is a constant vector. In this situation, by defining the energy of the shifted solution $\phi(\vec{x} + \vec{k})$ as $E_{\vec{k}}$, we have found that, to minimize the energy for $\vec{k}=0$, one must impose the condition \eqref{cond} and, also, take $\det({\cal H})>0$ and at least one of the diagonal components of ${\cal H}$ to be positive, where 
\be\label{stabtrans}
{\cal H}_{ij}=-\int_{\mathbb{R}^D} d^Dx\,\LL_{x^i x^j}
\ee
represents the components of the Hessian matrix ${\cal H}$.

Since we have found that the Lagrangian density \eqref{lagrangian} may support a first-order formalism which relies on the null divergence of $\tensor{\Tc}{^\mu_\nu}$, let us now investigate it considering that the field and impurity are static. In this case, Eq.~\eqref{fogeral} allows us to take $\sigma^0 = G^0=0$, leading us with the vectors $\vec{\sigma} = (\sigma^1,\sigma^2,\ldots,\sigma^{D})$ and $\vec{G} = (G^1,G^2,\ldots,G^{D})$. The equation of motion \eqref{eoml} reads
\be\label{eomlstatic}
\nabla\cdot(F_X\left(\nabla\phi-\vec{\sigma}\right)) = -F_\phi +\vec{\sigma}\cdot\vec{G}_\phi,
\ee
where the dynamical term becomes simpler, in the form $X=-\frac12\left|\nabla\phi - \vec{\sigma}\right|^2$. The components \eqref{emtcomp} of $\tensor{\Tc}{^\mu_\nu}$ are now given by
\bes\label{emtcompstatic}
\begin{align}\label{t00static}
\tensor{\Tc}{^0_0}&=- F + \vec{\sigma}\cdot\vec{G},\\
\tensor{\Tc}{^0_i}&=\tensor{\Tc}{^i_0}=0,\\ \label{tijstatic}
\tensor{\Tc}{^i_j}& = F_X\left(\partial^i\phi + \sigma^i\right)\partial_j\phi - \delta^i_ j\left(F - \vec{\sigma}\cdot\vec{G}\right).
\end{align}
\ees

To obtain a first-order equation in the static case, we must write $X$ in terms of $\phi$. This can be done using the static version of Eq.~\eqref{XG}, which reads
\be\label{XGstatic}
-2XF_X^2 = \big|\vec{G}\big|^2,
\ee
where $\big|\vec{G}\big| = \sqrt{\vec{G}\cdot\vec{G}}$. If one can obtain a expression in the form $X=H(\phi,\vec{G})$ from the above equation, Eq.~\eqref{fogeral} leads us to the following first-order equation
\be\label{fostatic}
\nabla\phi=\vec{\sigma}+ \frac{\vec{G}}{F_X\big|_{X=H\left(\phi,\vec{G}\right)}}.
\ee
Notice that the above expression is, actually, a set of $D$ partial differential first-order equations. It is a condition to ensure the null divergence of the tensor $\tensor{\Tc}{^\mu_\nu}$. In order to get compatibility of Eq.~\eqref{fostatic} with the equations of motion \eqref{eomlstatic}, we must impose the constraint
\be\label{conststatic}
\frac{\vec{G}\cdot\vec{G}_\phi}{F_X}=-F_\phi,
\ee
as expected from Eq.~\eqref{constraint}.

To ensure stability under contractions and dilations, we impose Eqs.~\eqref{derrick1}, which leads us to
\be\label{derrickF1}
\int_{\mathbb{R}^D}d^Dx\,\left(D\left(F - \vec{\sigma}\cdot\vec{G}\right) +\vec{G}\cdot\nabla\phi \right)=0.
\ee
Let us now suppose that the energy density in Eq.~\eqref{t00static} can be written as a divergence, by taking
\be\label{divw}
- F + \vec{\sigma}\cdot\vec{G} = \nabla\cdot\vec{W},
\ee
in which $\vec{W} = (W^1,W^2,\ldots,W^D)$ depends only on $\phi$, such that the energy \eqref{energy} reads
\be\label{energyw}
E = \int_{\mathbb{R}^D}d^Dx\,\nabla\cdot \vec{W},
\ee
where $\vec{W}$ must be chosen to lead to localized structures with finite energy. We highlight that not all the functions which solve the constraint in Eq.~\eqref{constraint} can be used, as a constant of integration usually appears and leads to infinite energy. Considering Eq.~\eqref{divw}, we get from Eq.~\eqref{tijstatic} that $\tensor{\Tc}{^i_j} = -G^i\partial_j\phi + \delta^i_j \,\nabla\cdot\vec{W}$, which leads us to $
\tensor{\Tc}{^i_i} =(D\vec{W}_\phi-\vec{G})\cdot\nabla\phi$. Therefore, we see that, to get localized $\tensor{\Tc}{^i_i}$, both $\vec{G}$ and $\vec{W}_\phi$ must vanish asymptotically. We then take
\be\label{gw}
\vec{G}=\vec{W}_\phi,
\ee
which leads to $\tensor{\Tc}{^i_i} = (D-1)\,\nabla\cdot\vec{W}$. By combining this with Eqs.~\eqref{derrickF1} and the energy density \eqref{divw} we obtain the condition
\be\label{condenergy}
(D-1)\int_{\mathbb{R}^D}d^Dx\,\nabla\cdot\vec{W}=0.
\ee
This is required to get solutions which are stable against rescale of argument. Since the energy is given by Eq.~\eqref{energyw}, the above expression is equivalent to $(D-1)E=0$. This shows that the energy must be null for $D>1$. The case $D=1$ is special, as the above equation becomes an identity, such that the energy may have any value, depending on the specific model under investigation. These consequences are compatible with the results obtained in Refs.~\cite{imp10,imp11,imp12,imp13,imp17} and are related to the choice in Eq.~\eqref{gw}. Moreover, from the expression below Eq.~\eqref{gw}, one can see that the stress is null for any solution in $D=1$. For arbitrary dimensions, one can take advantage of the equations \eqref{fostatic} and \eqref{divw} to write the energy density \eqref{t00static} as
\be\label{rhonablaphi}
\tensor{\Tc}{^0_0} = F_X\left(\nabla\phi -\vec{\sigma}\right)\cdot\nabla\phi,
\ee
in which the eventual dependences on $\phi$ that may arise due to the presence of $F_X$ must be substituted in terms of $\nabla\phi$ and $\vec{\sigma}$ with the use of Eq.~\eqref{XGstatic}. There is another equivalent expression, given by
\be\label{rhophi}
\tensor{\Tc}{^0_0}= \left(\frac{1}{F_X}\,\vec{W}_\phi + \vec{\sigma}\right)\cdot\vec{W}_\phi.    
\ee
In this situation, one must use Eq.~\eqref{XGstatic} to write the above equation explicitly in terms $\phi$ and $\vec{\sigma}$.

The stability against contractions and dilations require not only Eq.~\eqref{condenergy}, but also Eq.~\eqref{derrick1b}, which reads
\be\label{condderrickf2}
\begin{aligned}
 \int_{\mathbb{R}^D} d^Dx\bigg\{&F_X \big(\nabla\phi+(\vec{x}\cdot\nabla)\,\vec{\sigma}\big)^2\\
    &-F_{XX}\Big[\big(\nabla\phi-\vec{\sigma}\big)\cdot\big(\nabla\phi+(\vec{x}\cdot\nabla)\,\vec{\sigma}\big)\Big]^2\bigg\}>0,
\end{aligned}
\ee
where we have used Eqs.~\eqref{fostatic}, \eqref{gw} and \eqref{condenergy}. The notation $\vec{x}=(x^1,x^2,\ldots,x^D)$ was used. This condition ensures that the solution $\phi(\vec{x})$ is stable against rescale of argument. A way to satisfy it is by considering models with $F_X>0$ and $F_{XX}\leq0$. There are, however, other possibilities; one of them will be presented in Sec.~\ref{secbi}.

We are also interested in the stability under spatial translations, as our model now does not support translational invariance. As we have commented right above Eq.~\eqref{stabtrans}, it is ensured by the condition \eqref{fostatic}, necessary to get the null divergence of $\tensor{\Tc}{^\mu_\nu}$, and also by $\det({\cal H})>0$ and by requiring that at least one of the diagonal terms of ${\cal H}$ is positive, where the components of the Hessian matrix ${\cal H}$ in Eq.~\eqref{stabtrans} are now
\be\label{stabtransf}
{\cal H}_{ij} = \int_{\mathbb{R}^D}d^Dx\,F_X\Mc_{kl}\partial_i\sigma^k\partial_j\sigma^l,
\ee
where we have defined the matrix $\Mc$, such that
\be\label{mijgeral}
\Mc_{ij} = \delta_{ij}-\frac{F_{XX}}{F_X}\,\left(\partial_i\phi+\sigma_i\right)\left(\partial_j\phi+\sigma_j\right).
\ee

We then investigate the behavior of a static solution $\phi(\vec{x})$ of Eq.~\eqref{eomlstatic} around small fluctuations $\eta(x^\mu)$. By taking the field as $\phi(x^\mu) = \phi(\vec{x}) + \eta(x^\mu)$ in Eq.~\eqref{eoml}, we get that the contributions in powers up to first-order in $\eta$ lead to
\be\label{stabgen}
\begin{aligned}
    \partial_\mu\left((F_X \delta^{\mu}_{\nu} +F_{XX}(\partial^\mu\phi + \sigma^\mu)(\partial_\nu\phi+\sigma_\nu))\partial^\nu\eta\right) = \\\left(-\partial_\mu\left(F_{X\phi}(\partial^\mu\phi + \sigma^\mu)\right)+F_{\phi\phi}+\sigma_\mu G^\mu_{\phi\phi}\right)\eta.
\end{aligned}
\ee
Considering that the fluctuations have the form $\eta(\vec{x},t)=\sum_k\,\eta_k(\vec{x})\cos(\omega_k t)$, we get
\be
\begin{aligned}
-\partial_i\!\left(F_X\partial_i\eta_k\right) -\partial_i\!\left(F_{XX}(-\partial_i\phi + \sigma^i)(\partial_j\phi-\sigma^j)\partial_j\eta_k\right) = \\\left(\partial_i\!\left(F_{X\phi}(\partial_i\phi - \sigma^i)\right)+F_{\phi\phi}-\sigma^i G^i_{\phi\phi}\right)\eta_k+ \omega_k^2F_X\eta_k.
\end{aligned}
\ee
This expression can be written in the form
\be\label{stabstatic}
\begin{aligned}
    -\frac{1}{F_X}\partial_i\!\left(F_X\Mc_{ij}\partial_j\eta_k\right) + U(\vec{x})\eta_k=\omega^2_k\eta_k,
\end{aligned}
\ee
where $\Mc_{ij}$ is given by Eq.~\eqref{mijgeral} and
\be\label{stabpot}
U(\vec{x}) = -\frac{\partial_i\!\left(F_{X\phi}\!\left(\partial_i\phi+\sigma_i\right)\right) + F_{\phi\phi} - \sigma^i G^i_{\phi\phi}}{F_X}
\ee
represents the stability potential. The expression in \eqref{stabstatic} is a partial differential equation in arbitrary dimensions whose solutions determine the allowed eigenfunctions and eigenvalues. For more on this issue, see Refs.~\cite{pde1,pde2}. It may also be written in the form $L\eta_k=\omega_k^2\eta_k$, where the operator $L$ is
\be\label{Lop}
L = -\frac{1}{F_X}\,\partial_i F_X \Mc_{ij}\partial_j +U(\vec{x}).
\ee
If the first-order equation \eqref{fostatic} is considered, the above expression can be factorized in terms of adjoint operators, in the form $L=S_i^\dagger S_i$, where
\bes\label{ssdaggergeral}\bal
& S_i=\Ac_{ij}\left(-\partial_j + P_j\right),\\
& S_i^\dagger = \Ac^\dagger_{ij}\left(\partial_j + P_j + {\Ac^\dagger}^{-1}_{jk} F_X^{-1}\partial_j\left(F_X \Ac^\dagger_{kj}\right)\right),
\eal
\ees
in which $\Ac_{ij}$ represents the coordinate-dependent components of the matrix $\Ac$. It must obey $\Mc_{ij}= \Ac^\dagger_{ki}\Ac_{kj}$, with $\Mc_{ij}$ as in Eq.~\eqref{mijgeral}. Also, $\Ac^\dagger$ is the conjugate transpose of $\Ac$ and ${\Ac^\dagger}^{-1}$ is the inverse of $\Ac^\dagger$. Unfortunately, finding the components of the matrix $\Ac$ is not an easy task. However, we have found that, in three spatial dimensions ($D=3$), they can be written in the form 
\be
\Ac_{ij}=\Ac_{ij}^\dagger=\delta_{ij}\sqrt{1+\alpha^2\,(\partial_i\phi+\sigma_i)^2} +{\bf i}\,\alpha\,\epsilon_{ijk}(\partial_k\phi+\sigma_k),
\ee
where ${\bf i}=\sqrt{-1}$ and $\alpha^2=-F_{XX}/F_X$. The vector function $P_i$ is defined through the equation $P_i\partial_j\phi=\partial_j\!\left(\partial_i\phi+\sigma_i\right)$. By using Eqs.~\eqref{fostatic} and \eqref{gw}, it can be obtained explicitly, in the form
\be
P_i = \frac{d}{d\phi}\left(\frac{W_\phi^i}{F_X\big|_{X=H\left(\phi,\vec{W}_\phi\right)}}\right),
\ee
 Therefore, the operators $S_i$ and $S^\dagger_i$ are defined in terms of the static solutions $\phi(\vec{x})$ of the first-order equation \eqref{fostatic}. The zero mode, must obey the equation $S_i\eta_0=0$, or $\partial_i\eta_0=P_i\eta_0$. Notice that this defines a set of partial differential equations, so we cannot ensure that the zero mode will exist in general for Lagrangian densities with the form \eqref{lagrangian} in arbitrary dimensions. It is worth remarking that the factorization of the operator \eqref{Lop} in terms of the product of the adjoint operators \eqref{ssdaggergeral} does not ensure that negative eigenvalues are absent, as $S_i$ and $S^\dagger_i$ may engender divergences that lead to instability. Further analysis is required depending on the system under investigation.

\subsection{Single spatial dimension}\label{secstaticD1}
Even though our procedure allows for the presence of first-order equations in arbitrary dimensions, Eq.~\eqref{condenergy} shows us that the static case with a \textit{single} spatial dimension ($D=1$) is special, as the energy depends on the specific model under investigation. In this situation, both the impurity and the auxiliary function become scalars, so we denote them simply by $\sigma(x)$ and $G(\phi)$. The equation of motion \eqref{eomlstatic} now takes the form 
\be\label{eomlstatic1D}
\left(F_X\left(\phi'-\sigma\right)\right)' = -F_\phi +\sigma G_\phi,
\ee
where the prime indicates derivative with respect to $x$, and $X=-\frac12(\phi'-\sigma)^2$. Notice that the above equation is an ordinary differential equation. The first-order equation \eqref{fostatic} reads
\be\label{fostatic1D}
\phi^\prime=\sigma+ \frac{G}{F_X\big|_{X=H(\phi,G)}}.
\ee
To make this equation compatible with the equation of motion \eqref{eomlstatic1D}, we must consider the constraint \eqref{constraint}, which reads $GG_\phi/F_X=-F_\phi$. We also remark that in the case $D=1$, the above equation is not a set of partial differential equations as in the case of arbitrary dimensions. Instead, the first-order equation is just an ordinary (although nonlinear) differential equation.

As we have previously commented, the value of the energy in a single spatial dimension depends on the model under investigation. Indeed, the expression in Eq.~\eqref{energyw} simplifies to
\be\label{ewD1}
E = W(\phi(\infty))-W(\phi(-\infty)),
\ee
depending only on the auxiliary function calculated at the asymptotic values of the solutions of the first-order equation \eqref{eomlstatic1D}.

We then investigate the stability against contractions and dilations. Since \eqref{fostatic1D} makes the $\tensor{\Tc}{^\mu_\nu}$ having null divergence, one can show that $\tensor{\Tc}{^1_1}=C$, where $C$ is a constant. From the condition \eqref{derrick1a}, we get $C=0$, which leads to $F -2XF_X=0$. Notice that this equation is \emph{local}, whilst the one in Eq.~\eqref{derrick1a} for arbitrary dimensions is global. Moreover, if $F_X$ is positive, we must have $F$ negative to get stable solutions. The condition \eqref{derrick1b} simplifies to
\be\label{derrick1D}
 \int^{+\infty}_{-\infty} dx F_X A^2\big(\phi^\prime+x\sigma^\prime\big)^2>0,
\ee
where
\be\label{a2hyp}
A^2=1+2X\frac{F_{XX}}{F_X},
\ee
which is equal to $\Mc_{11}$ in Eq.~\eqref{mijgeral} for $D=1$. The stability under spatial translations requires that the first-order equation \eqref{fostatic1D} is used and that ${\cal H}_{11}$ from Eq.~\eqref{stabtrans} is positive. The latter condition becomes simpler, in the form
\be\label{derricktrans1D}
\int^{+\infty}_{-\infty} F_XA^2{\sigma^\prime}^2 dx>0.
\ee

Furthermore, for $D=1$, the equation \eqref{stabstatic} that describes the linear stability simplifies to an eigenvalue equation of the Sturm-Liouville type, 
\be\label{stabstaticx}
\begin{aligned}
&-\frac1{F_X}\left(A^2F_X\eta_k^\prime\right)^\prime + U(x)\eta_k = \omega_k^2\eta_k,
\end{aligned}
\ee
where $A^2$ is as in Eq.~\eqref{a2hyp} and
\be
U(x)=-\frac{\left(F_{X\phi}\left(\phi^\prime-\sigma\right)\right)^\prime+F_{\phi\phi}-\sigma G_{\phi\phi}}{F_X}.
\ee
To preserve the hyperbolicity of the stability equation \eqref{stabstaticx}, we impose that $A^2>0$ and $F_X>0$. Interestingly, these conditions make Eqs.~\eqref{derrick1D} and \eqref{derricktrans1D} become identities, which ensure stability against translation and rescale. The linear stability, however, is not guaranteed. Nevertheless, if the first-order equation \eqref{fostatic1D} is satisfied, we can factorize the Sturm-Liouville operator associated to the eigenvalue equation $L\eta_k=\omega^2_k\eta_k$ in Eq.~\eqref{stabstaticx} as a product of adjoint operators. We then write $L=S^\dagger S$, where
\bes
\bal
S &=A\left(-\frac{d}{dx} + \frac{\phi^{\prime\prime} - \sigma^\prime}{\phi^\prime}\right),\\
S^\dagger &=A\left(\frac{d}{dx} + \frac{\phi^{\prime\prime} - \sigma^\prime}{\phi^\prime} + \frac{(AF_X)^\prime}{AF_X}\right).
\eal
\ees
By using $S\eta_0=0$ we get that the zero mode is given by
\be\label{zeromode}
\eta_0(x) = \phi^\prime e^{-\int dx\frac{\sigma^\prime}{\phi^\prime}}.
\ee
We then see that the application of our formalism to the case of a single spatial dimension generalizes the results obtained in Ref.~\cite{stabkink}, where the linear stability of impurity-free models ($\sigma=0$) was investigated. Since the impurity is, in principle, arbitrary, we cannot be sure how the above zero mode will behave. Nevertheless, the linear stability is ensured if $\eta_0(x)$ does not engender nodes.

\section{Examples}\label{secexamples}

In the previous Sec. \ref{sec2} we have organized the general results of the present study. There, the model was defined and investigated, showing the possibility of finding first-order differential equations that solves the equation of motion on general grounds. This was achieved under the presence of a new tensor, $\tensor{\Tc}{^\mu_\nu}$, whose divergence can be made to vanish under the presence of a constraint. We have also considered the case of static solutions, including a full examination of stability under rescaling and translation of the spatial coordinates, and the addition of small fluctuations to the field configurations. Due to the importance of these results, in this section we illustrate the general findings investigating some specific possibilities, as the canonical model and two others, having  kinematic modifications of the $k$-field and Born-Infeld type.

\subsection{Canonical model}
Let us now consider the canonical Lagrangian coupled to impurities, such that
\be\label{canonical}
F(\phi,X) =  X - V(\phi).
\ee
 The impurity-free case, in which $\sigma_\mu=0$, recovers the standard case, $\LL = \frac12\partial_\mu\phi\partial^\mu\phi - V(\phi)$. The general aspects of this model were investigated in Ref.~\cite{letterimpurity}, where one finds the equation of motion, the first-order equation that arises from the condition \eqref{cond} and the potential which satisfies the constraint \eqref{constraint}. Here, we go straight to study of the stability of the solutions. To do so, we consider static field and impurity whose associated first-order equation obtained from \eqref{fostatic} with \eqref{gw} is
\be\label{focanonical}
\nabla\phi = \vec{\sigma} +\vec{W_\phi}.
\ee
It is compatible with the equation of motion \eqref{eomlstatic}
\be
\nabla^2\phi=\nabla\cdot\vec{\sigma} + \vec{W}_{\phi\phi}\cdot\vec{W}_{\phi}  +\vec{\sigma}\cdot\vec{W}_{\phi\phi},
\ee
if the potential has the form
\be\label{vwcanonical}
V(\phi) = \frac12\,\big|\vec{W}_\phi\big|^2,
\ee
where $\big|\vec{W}_\phi\big|=\sqrt{\vec{W}_\phi\cdot\vec{W}_\phi}$ and we have taken zero for the integration constant to avoid infinite energy as we have commented in the discussion below Eq.~\eqref{energyw}. The presence of $\vec{W}$ allows us to calculate the energy as in Eq.~\eqref{energyw}. We highlight that, as shown in Ref.~\cite{letterimpurity}, this model supports a BPS bound in the lines of Refs.~\cite{bogo,ps}. If the potential has the form \eqref{vwcanonical} and the first-order equation \eqref{focanonical} is satisfied, then the energy is minimized to the value given by the expression in Eq.~\eqref{energyw}. This confirms the stability of the solution. It is worth pointing out that our procedure recovers the results obtained in Refs.~\cite{imp10,imp11,imp12,imp13} for static field in a single spatial dimension, with the very same first-order equation and potential. In higher dimensions, our model generalizes Ref.~\cite{imp17}, since we can take the vector $\vec{W}_\phi$ with different components, to account for distinct self-interactions, one for each spatial direction.
 
First, to ensure stability against contractions and dilations, we must analyze Eq.~\eqref{condderrickf2}, which reads $\int_{\mathbb{R}^D} d^Dx (\nabla\phi+(\vec{x}\cdot\nabla)\,\vec{\sigma})^2>0$. This expression is an identity, so the stability against rescale is ensured in \emph{arbitrary dimensions} for solutions of \eqref{focanonical} obeying Eq.~\eqref{condenergy}. Second, we look into the stability against translations by using Eq.~\eqref{stabtransf}, which becomes
\be\label{stabtransstd}
{\cal H}_{ij} = \int_{\mathbb{R}^D}d^Dx\,\partial_i\vec{\sigma}\cdot\partial_j\vec{\sigma}.
\ee
This Hessian matrix must have positive determinant and at least one positive component in the diagonal in order to avoid instabilities. This condition must be analyzed for each impurity given. Third, and by last, to guarantee that the solution remains stable under small fluctuations, we analyze the linear stability. It is governed by Eq.~\eqref{stabstatic}, which reads 
\be\label{canonicalstabstatic}
-\nabla^2\eta_k +U(\vec{x})\eta_k= \omega_k^2\eta_k,
\ee
where the stability potential has the form
\be\label{ustd}
U(\vec{x}) =\big|\vec{W}_{\phi\phi}\big|^2+\vec{W}_{\phi}\cdot\vec{W}_{\phi\phi\phi}+\vec{\sigma}\cdot \vec{W}_{\phi\phi\phi},
\ee
with $\big|\vec{W}_{\phi\phi}\big|=\sqrt{\vec{W}_{\phi\phi}\cdot\vec{W}_{\phi\phi}}\,$. This equation can be written in the form $H\eta_k=\omega^2_k\eta_k$, where the Schr\"odinger-like operator $H$ is
\be
H = -\nabla^2 + U(\vec{x}).
\ee
If the field solves the first-order equation \eqref{focanonical}, this operator can be factorized in terms of adjoint operators, as $H=\vec{S}^\dagger\cdot \vec{S}$, where
\be
\vec{S} = -\nabla + \vec{W}_{\phi\phi},\qquad \vec{S}^\dagger = \nabla + \vec{W}_{\phi\phi}.
\ee
Supposing that these operators are smooth everywhere, the stability equation \eqref{canonicalstabstatic} does not support negative eigenvalues. This ensures that the solutions are stable against small fluctuations. The zero mode (when exists) obeys the equation $\vec{S}\eta_0=0$, which leads us to
\be\label{eta0std}
\nabla\eta_0 = \vec{W}_{\phi\phi}\,\eta_0.
\ee
This is a partial differential equation that not always admits solution, so the zero mode may not exist.

To illustrate the above results, let us investigate localized structures in two spatial dimensions. In this situation, both the impurity and $\vec{W}$ have two components. We then use the solution obtained in Ref.~\cite{letterimpurity} which arises for $\vec{W}=(W(\phi),W(\phi))$, where
\be\label{wpmodel}
W(\phi) = \frac{3}{5}\phi^{5/3} -\frac{3}{7}\phi^{7/3},
\ee
and for the impurity vector having the form
\be\label{sigmaD2std}
\vec{\sigma}(x,y)=(6x - 1,6y - 1)\tanh^2(x^2 + y^2)\,\sech^2(x^2 + y^2).
\ee
The first-order equation \eqref{focanonical} admits the solution
\be\label{solD2std}
\phi(x,y) = \tanh^3(x^2+y^2).
\ee
We have shown that, regardless the impurity, the stability against rescale of the solutions is ensured. However, we still have to deal with the stabilities with respect to translations and small fluctuations. To ensure that our solution has the energy minimized at its location, i.e., there is no other point of the space with less energy, we must use the Hessian matrix in \eqref{stabtransstd}. This calculation cannot be done analytically for the impurity \eqref{sigmaD2std}. By using numerical procedures, we have shown that its determinant is positive. Therefore, the solution \eqref{solD2std} is stable against translations in the space.

\begin{figure}[t!]
    \centering
\includegraphics[width=0.8\linewidth,trim={1.5cm 4cm 0cm 0cm},clip]{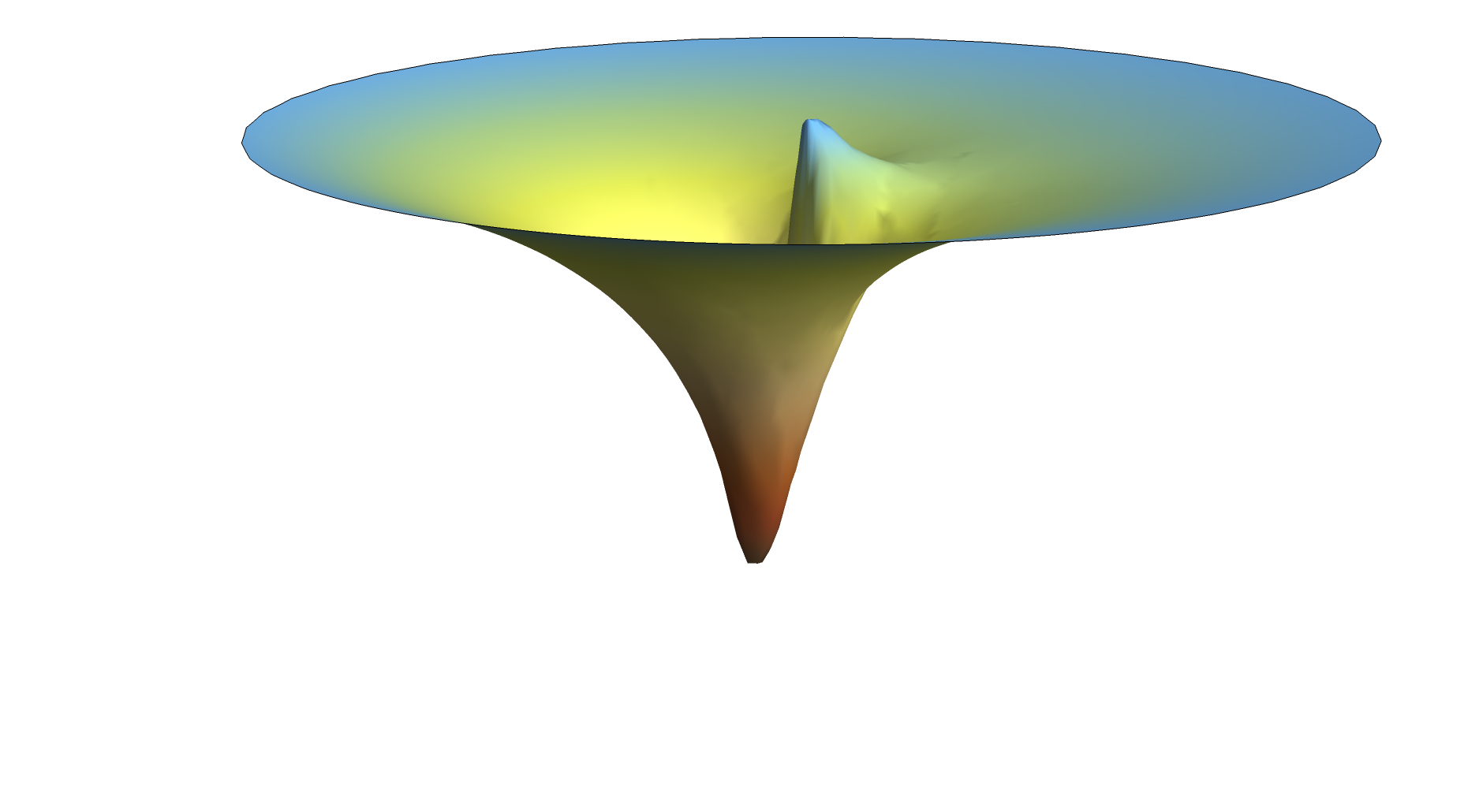}
    \caption{The lowest bound state allowed by the stability equation \eqref{canonicalstabstatic} with the potential \eqref{ustdexD2}, whose eigenvalue is $\omega^2\approx0.884$.}
    \label{modesstdD2}
\end{figure}

We then turn our attention to the linear stability, governed by Eq.~\eqref{canonicalstabstatic}. The stability potential \eqref{ustd} can be written as
\be\label{ustdexD2}
\begin{aligned}
    U(x,y)&=\frac{8-4\,(9x+9y+8)\,\sech^2(x^2+y^2)}{9\tanh^2(x^2+y^2)} \\
&+\frac{8\,(3x+3y+4)\,\sech^4(x^2+y^2)}{9\tanh^2(x^2+y^2)}.
\end{aligned}
\ee
This potential diverges at the origin and has a minimum at $(x,y)=(0.223,0.223)$, such that $U(0.223,0.223)=-6.031$. Asymptotically, we have $U(x,y)\approx8/9 + 16(x+y)e^{-2(x^2+y^2)}$, so it tends to be a constant far away from the origin. Therefore, the possible eigenvalues associated to the bound states must be in the interval $[-6.031,8/9]$. The zero mode must be calculated from Eq.~\eqref{eta0std}, which leads us to the set of equations
\be\label{eta0std1}
\frac{\partial\eta_0}{\partial x} = \frac{\partial\eta_0}{\partial y} = W_{\phi\phi}\,\eta_0.
\ee
This requires that $\eta_0=\eta_0(\xi)$, where $\xi=x+y$. So, the above set of equations can be written in the form $d\eta_0/d\xi = W_{\phi\phi}\eta_0(\xi)$. Since $W_{\phi\phi}$ does not depend exclusively on $\xi$, this equation does not admit solutions. Therefore, the zero mode does not exist in this model. This conclusion is valid not only for the auxiliary function in \eqref{wpmodel}, but also to any $W(\phi)$ such that $\vec{W}=(W(\phi),W(\phi))$. In Fig.~\ref{modesstdD2}, we have displayed the lowest bound state allowed by the stability equation \eqref{canonicalstabstatic} with the potential \eqref{ustdexD2}, whose eigenvalue is $\omega^2\approx0.884$. The zero mode is absent and only positive eigenvalues are allowed, ensuring that the solution \eqref{solD2std} is linearly stable.

\subsubsection{Single spatial dimension}
Next, we investigate solutions of the first-order equation \eqref{focanonical} in $D=1$, with the auxiliary function and impurity given by the same form studied in Ref.~\cite{letterimpurity},
\bal
& W(\phi) = \phi-\frac13\phi^3,\\ \label{sigma1}
&\sigma(x)=\frac{\left(\sqrt{1+a}-\sqrt{1+a\tanh^2(x)}\right)\sech^2(x)}{\left(1+a\tanh^2(x)\right)^{3/2}},
\eal
where $a$ is a real parameter that obeys $a>-1$. The expression in Eq.~\eqref{vwcanonical} combined with the above function $W(\phi)$ leads us to the well-known $\phi^4$ model, i.e., $V(\phi) = (1-\phi^2)^2/2$. The first-order equation \eqref{focanonical} combined with the above functions supports the solution
\be\label{solD1}
\phi(x) = \frac{\sqrt{1+a}\,\tanh(x)}{\sqrt{1+a\tanh^2(x)}},
\ee
where we have used the condition $\phi(0)=0$ to determine the constant of integration that arises in the process. The study of the linear stability depends on the potential \eqref{ustd}, which reads
\be\label{ustdD1}
U(x) = \frac{4(1+a)\tanh^2(x)}{1+a\tanh^2(x)} -\frac{2\sqrt{1+a}\,\sech^2(x)}{\left(1+a\tanh^2(x)\right)^{3/2}}.
\ee
The zero mode in Eq.~\eqref{zeromode} is
\be\label{etazeroD1}
\eta_0(x) =  \frac{{\cal N}_a\,\sech^2(x)}{2\sqrt{(1+a)(1+aT^2(x))} + a(1+T^2(x)) +2},
\ee
where ${\cal N}_a = a^2\sqrt{2}/\sqrt{a(6+a)-2(2a+3)\ln(a+1)}$ is a constant of normalization and $T(x)=\tanh^2(x)$.

\begin{figure}[t!]
    \centering
\includegraphics[width=0.5\linewidth]{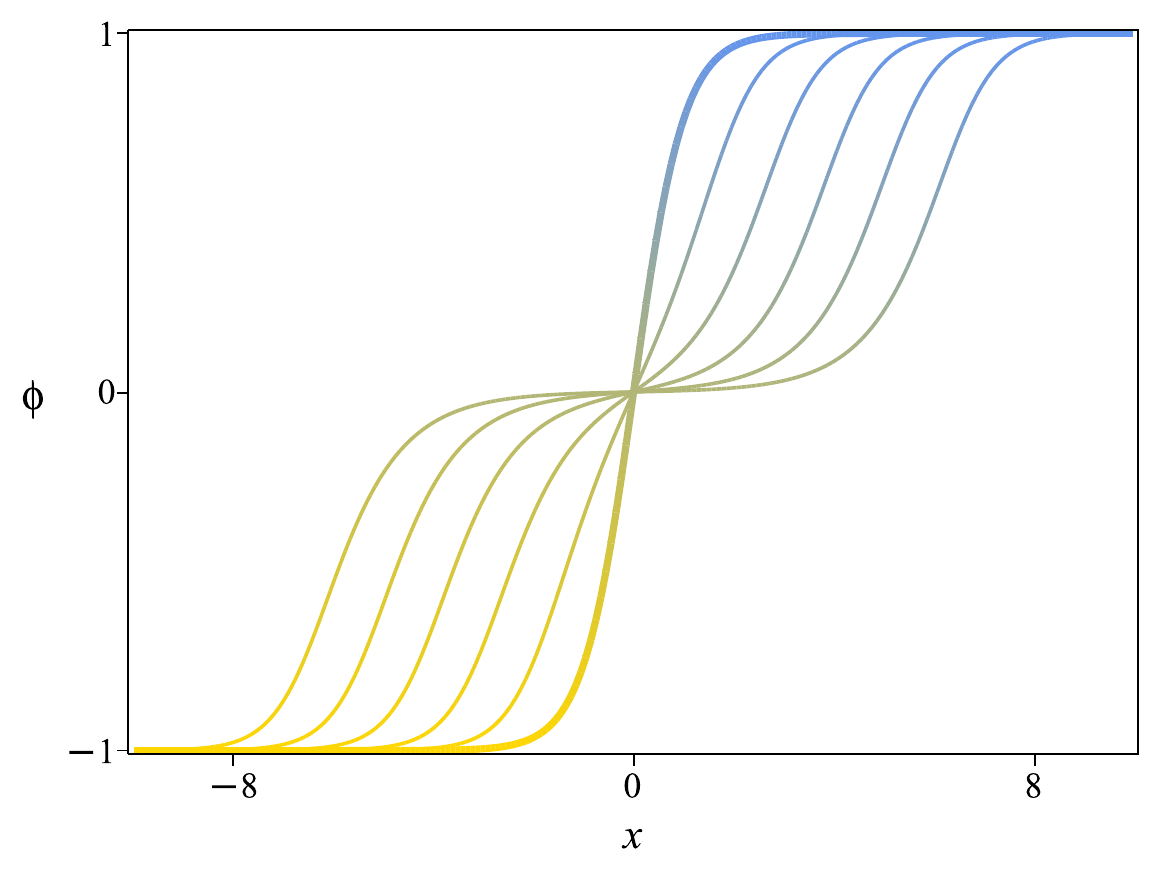}\includegraphics[width=0.5\linewidth]{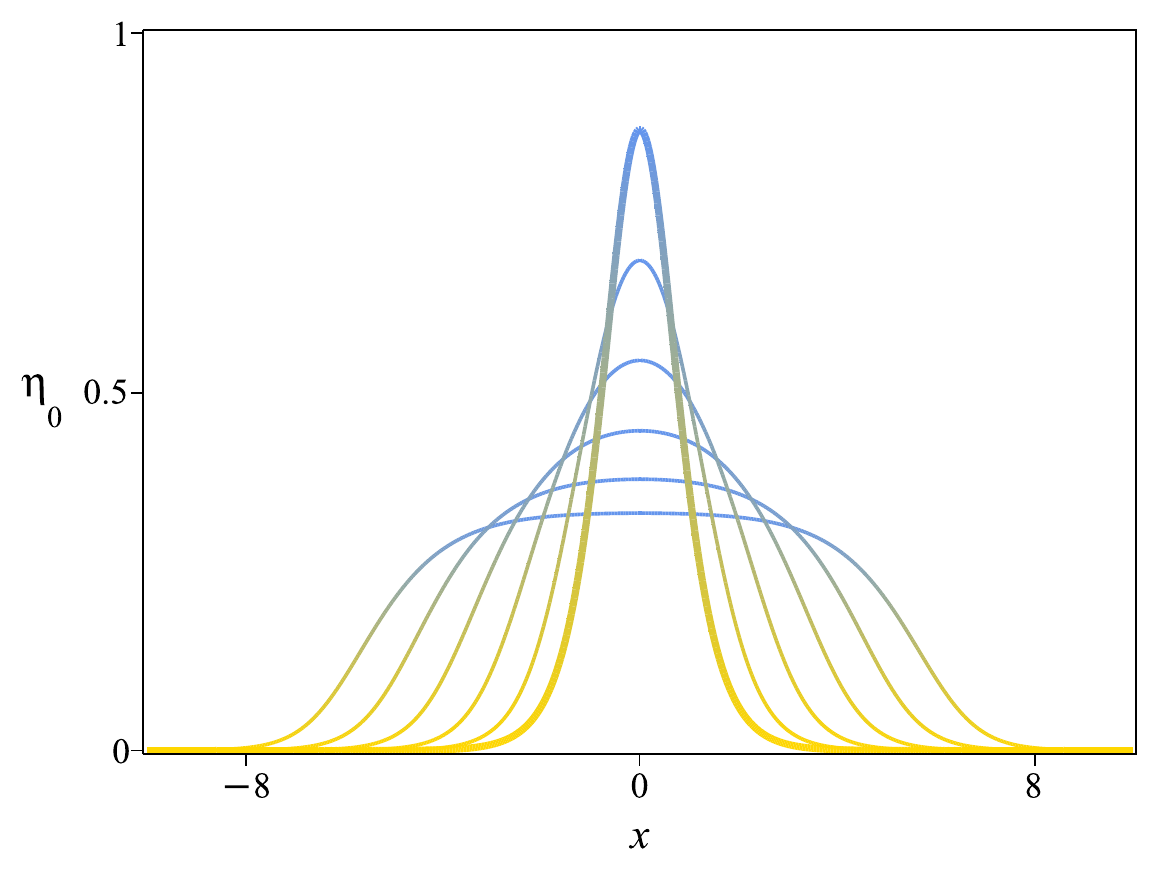}
    \caption{The solution \eqref{solD1} (left) and the zero mode \eqref{etazeroD1} (right) for $a=-0.99999,-0.9999,-0.999,-0.99,-0.9$ and $0$. The thickest line represents the case $a=0$ in each panel.}
    \label{figD1std}
\end{figure}

In Fig.~\ref{figD1std}, we display the solution \eqref{solD1} and the above zero mode for some values of $a$. We have only taken $a\leq0$ because positive values of $a$ only modify the thickness of the solution. Notice that, as $a$ approaches $-1$, the solution tends to get an inflection point with null derivative at the origin and the peak of the zero mode goes down. As expected from the general investigation of the stability, the zero mode does not present nodes, regardless the value of $a$; this confirms that the solution is linearly stable.

\subsection{$k$-field}
We then apply our formalism to generalized models. In this context, we may consider the so-called $k$-defects, in which the kinetic term is generalized to a function of the standard one. In particular, one can consider a power law of $X$, as introduced in the study of topological structures in Ref.~\cite{babichev}, and later investigated in Refs.~\cite{adamk,trilogia2,superk,zhongk}. Inspired in these works, we take
\be\label{fk}
F(\phi,X) = \frac{2^{n-1}}{n}\,X|X|^{n-1}-V(\phi),
\ee
where $n$ is a positive integer number. In this situation, Eq.~\eqref{momentuml} leads us to $\Pi^\mu = |2X|^{n-1}\left(\partial^\mu\phi + \sigma^\mu\right)$. Therefore, the equation of motion \eqref{eoml} for the case in Eq.~\eqref{fk} can be written in the form
\be\label{eomstd}
\partial_\mu\Pi^\mu = \sigma_\mu G^\mu_\phi - V_\phi.
\ee
To obtain a divergenceless $\tensor{\Tc}{^\mu_\nu}$, we take Eq.~\eqref{fogeral}. However, to write it, we must relate $X$ and $\phi$ via Eq.~\eqref{XG}, in the form $-X=\frac12 (-G_\mu G^\mu)^{1/(2n-1)}$. By substituting into Eq.~\eqref{fogeral}, we have
\be
\partial_\mu\phi = - \sigma_\mu -\frac{G_\mu}{|G_\alpha G^\alpha|^{(n-1)/(2n-1)}}.
\ee
By combining this with the constraint \eqref{constraint}, we obtain
\be
V(\phi) = \frac{2n-1}{2n}\left(-G_\mu G^\mu\right)^{n/(2n-1)} + C.
\ee
As before, we impose $G_\mu$ to be a spacelike vector and $C$ is an integration constant.

\subsubsection{Static case}
Let us now focus on static configurations in the model \eqref{fk}. From Sec.~\ref{staticgeneral}, we obtain that the first-order equation \eqref{fostatic} combined with \eqref{gw} takes the form
\be\label{fokstatic}
\nabla\phi = \vec{\sigma} + \frac{\vec{W}_\phi}{\big|\vec{W}_\phi\big|^{2(n-1)/(2n-1)}},
\ee
where we have used $X=-\frac12\,\big|\vec{W}_\phi\big|^{2/(2n-1)}$ from Eq.~\eqref{XGstatic} and $\big|\vec{W}_\phi\big|$ is as given below Eq.~\eqref{vwcanonical}. The above equation is compatible with the equation of motion if the potential is written as
\be\label{vwk}
V(\phi) = \frac{2n-1}{2n}\big|\vec{W}_\phi\big|^{2n/(2n-1)},
\ee
where we have taken the integration constant to be zero in order to avoid infinite energy, similarly as in the previous section. The energy density obtained from Eqs.~\eqref{rhonablaphi} and \eqref{rhophi} take the form
\be\label{t00kstatic}
\begin{split}
    \tensor{\Tc}{^0_0} &= |2X|^{n-1}\left(\nabla\phi -\vec{\sigma}\right)\cdot\nabla\phi\\
    &= \big|\vec{W}_\phi\big|^{2n/(2n-1)} + \vec{\sigma}\cdot\vec{W}_\phi.
\end{split}
\ee

Since we got the equations describing $k$-fields, we turn our attention to the stability of the static solutions. First, we look into the stability against contractions and dilations using Eq.~\eqref{condderrickf2}, which one can show is an identity because $F_X>0$ and $F_{XX}\leq0$ in this case, so the stability against rescale is ensured in \emph{arbitrary dimensions} for the generalized model \eqref{fk}. In this case, the components in Eq.~\eqref{mijgeral} becomes
\be\label{mijk}
\Mc_{ij}=\delta_{ij}+2(n-1)\frac{W_\phi^i W_\phi^j}{\big|\vec{W}_\phi\big|^2}.
\ee
Second, we look into the stability against translations by using Eq.~\eqref{stabtransf}, which reads
\be\label{stabtransk}
\begin{aligned}
    {\cal H}_{ij} = \int_{\mathbb{R}^D}d^Dx\,\big|\vec{W}_\phi\big|^{2(n-1)/(2n-1)}\partial_i\vec{\sigma}\cdot \partial_j\vec{\sigma} \\
    +2(n-1)\int_{\mathbb{R}^D}d^Dx\,\frac{W_\phi^k W_\phi^l}{|{\vec{W}_\phi}|^{2n/(2n-1)}}\partial_i\sigma^k\partial_j\sigma^l,
\end{aligned}
\ee
This Hessian matrix must have positive determinant and at least one positive component in the diagonal in order to avoid instabilities. The linear stability is governed by Eq.~\eqref{stabstatic}, whose stability potential \eqref{stabpot} now takes the form
\be\label{ukstatic}
\begin{aligned}
    U(\vec{x})&=\frac{2-2n}{2n-1}\,\big|\vec{W}_\phi\big|^\frac{6-8n}{2n-1}\left(\vec{W}_\phi\cdot \vec{W}_{\phi\phi}\right)^2\\
    &+\big|\vec{W}_\phi\big|^\frac{4-4n}{2n-1}\left(\big|\vec{W}_{\phi\phi}\big|^2 + \vec{W}_{\phi}\cdot \vec{W}_{\phi\phi\phi}\right)\\
    &+\big|\vec{W}_\phi\big|^\frac{2-2n}{2n-1}\, \vec{\sigma}\cdot\vec{W}_{\phi\phi\phi}.
\end{aligned}
\ee

To illustrate our procedure in this case, we show specific models in two and one spatial dimensions. For $D=2$, we consider the impurity given by Eq.~\eqref{sigmaD2std} and $\vec{W}(\phi)=(W(\phi),W(\phi))$, with
\be
\begin{aligned}
    W(\phi) &= \frac{3\cdot2^{n-1}}{4n+1}\,\phi^{(4n+1)/3} \\
   &\cdot\,_{2}F_{1}\!\left(1-2n,2n+\frac12;2n+\frac32;\phi^{2/3}\right)\!,
\end{aligned}
\ee
where the $_2F_1(a,b;c;z)$ represents the hypergeometric function of argument $z$. The potential is given by Eq.~\eqref{vwk}, which can be written as
\be
V(\phi) = \frac{2^{n-1}\,(2n-1)}{n}\,\phi^{2n}\left(\phi^{1/3}-\phi^{-1/3}\right)^{2n}.
\ee
This potential engenders minima at $\phi=\pm1$ and $\phi=0$ independently of $n$. In this situation, the first-order equation \eqref{fokstatic} admits the very same solution in Eq.~\eqref{solD2std}. The energy density can be calculated from Eq.~\eqref{t00kstatic}, which takes the form
\be
\tensor{\Tc}{^0_0}(x,y) = 3\cdot2^n(x+y)\tanh^{4n}(x^2 + y^2)\,\sech^{4n}(x^2 + y^2).
\ee
The behavior of the above energy density is similar to the one obtained in Ref.~\cite{letterimpurity}. The parameter $n$ only changes its thickness, so we do not display the behavior of the above expression here.

To analyze the stability, we must consider the Hessian matrix in Eq.~\eqref{stabtransk} for the impurity \eqref{sigmaD2std} and the solution \eqref{solD2std}. The calculation is numeric, but we have checked that its determinant is positive and ${\cal H}_{11}>0$ for $n\leq 50$, ensuring stability against spatial translations. We are also concerned with stability under small fluctuations, which is described by Eq.~\eqref{stabstatic} with \eqref{gw} and \eqref{mijk}. The stability potential is given by \eqref{ukstatic}, which can be written as
\be\label{un2D2}
\begin{aligned}
    U(x,y)&=\frac{8(2n-1)(6(n-1)(x+y)+1)}{9\tanh^2(x^2+y^2)}\\ 
    &-\frac{4(2n-1)(3(16n-13)(x+y)+8)\,S^2(x,y)}{9\tanh^2(x^2+y^2)} \\
&+\frac{8(2n-1)(3(8n-7)(x+y)+4)\,S^4(x,y)}{9\tanh^2(x^2+y^2)},
\end{aligned}
\ee
where have used the notation $S(x,y)=\sech(x^2+y^2)$. In this situation, we must be careful, as the behavior of the stability potential is $U(x,y)\propto(x+y)$ for $n\geq2$. In addition, all the components of the matrix $\Mc_{ij}$ in \eqref{mijk} are non-negative. Since the potential ranges from $-\infty$ to $+\infty$, there is a continuum of negative states, which leads to instability.

\subsubsection{Single spatial dimension}
\begin{figure}[t!]
    \centering
\includegraphics[width=0.5\linewidth]{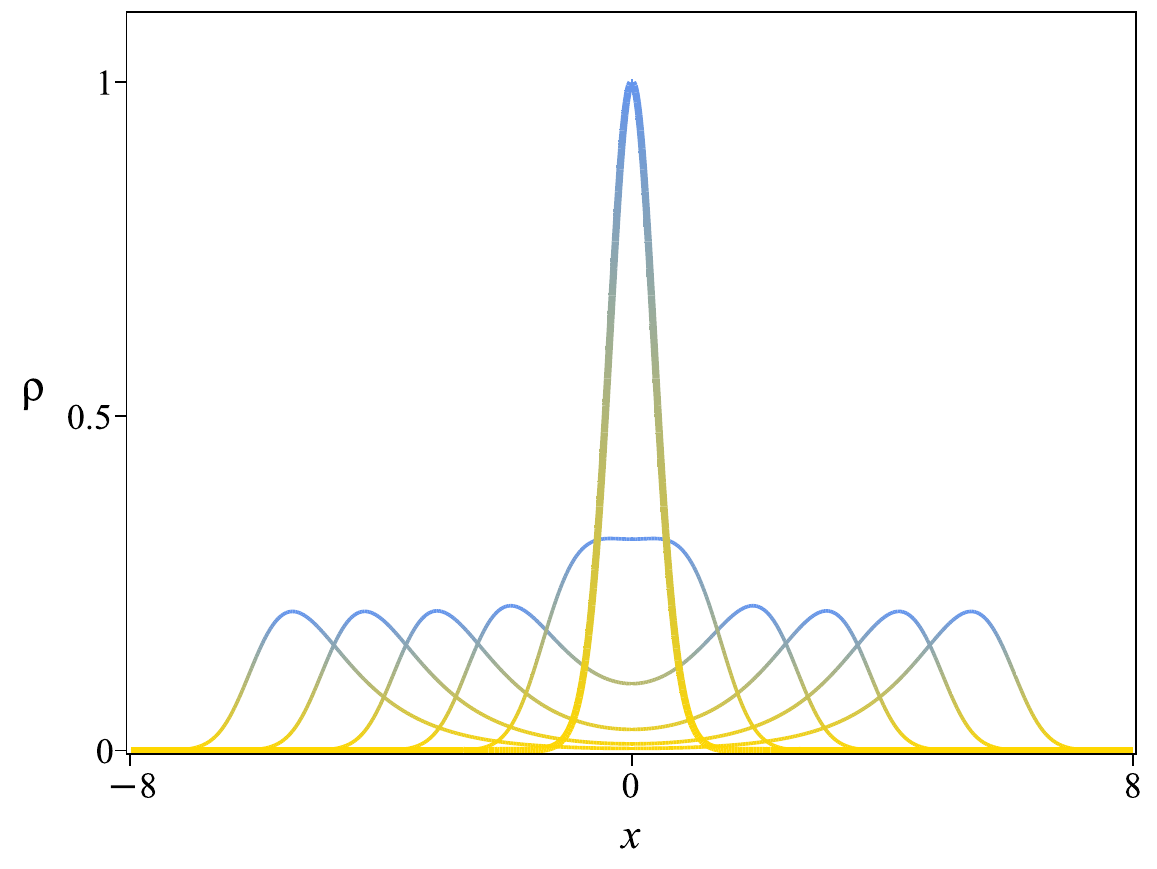}\includegraphics[width=0.5\linewidth]{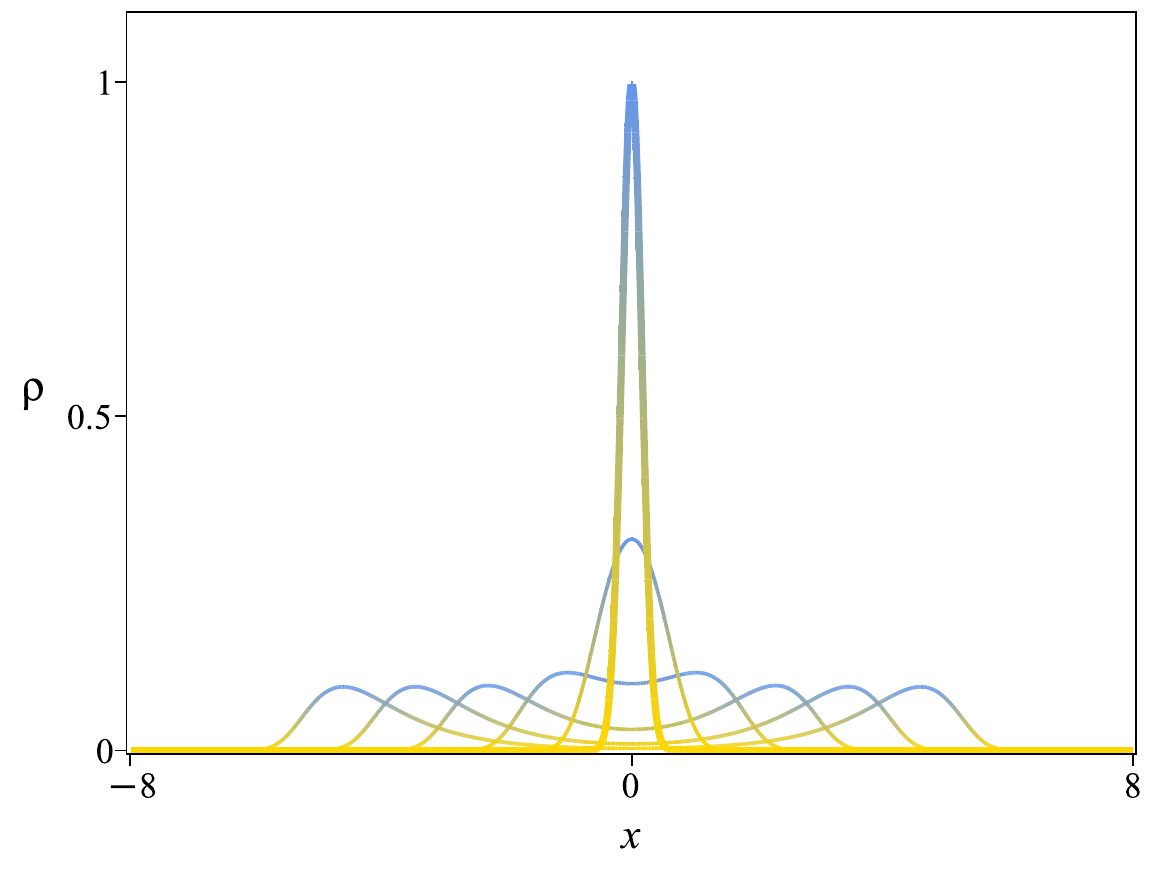}
    \caption{The energy density $\rho=\tensor{\Tc}{^0_0}$ in \eqref{t00D1k} with $a=-0.99999,-0.9999,-0.999,-0.99,-0.9$ and $0$, for $n=2$ (left) and $10$ (right). In both panels, the thickest line represents the case $a=0$.}
    \label{figD1k}
\end{figure}

In a single spatial dimension ($D=1$), both the first-order equation \eqref{fokstatic} and the energy density \eqref{t00kstatic} become simpler, in the form
\be
\phi^\prime = \sigma + \frac{W_\phi}{|W_\phi|^{2(n-1)/(2n-1)}}
\ee
and
\be
\begin{split}
    \tensor{\Tc}{^0_0} &= |2X|^{n-1}\left(\phi^\prime -\sigma\right)\phi^\prime\\
    &= |W_\phi|^{2n/(2n-1)} + \sigma W_\phi.
\end{split}
\ee
The only remaining component in Eq.~\eqref{mijk} is $\Mc_{11}=2n+1$, which is positive. This ensures the hyperbolicity of the Sturm-Liouville eigenvalue operator \eqref{stabstatic}, whose potential is now given by
\be
\begin{aligned}
    U(x)&=\frac{1}{2n-1}\,W_\phi^\frac{4(1-n)}{2n-1}W_{\phi\phi}^2\\
    &+W_\phi^\frac{3-2n}{2n-1}W_{\phi\phi\phi}+W_\phi^\frac{2(1-n)}{2n-1} \sigma W_{\phi\phi\phi}.
\end{aligned}
\ee
To illustrate this case, we take the auxiliary function
\be
W(\phi) = \phi\; {}_2F_1\left(\frac{1}{2},-2 n +1;\frac{3}{2};\phi^{2}\right),
\ee
where ${}_2F_1\left(a,b;c;z\right)$ is the hypergeometric function of argument $z$, and the same impurity as given in Eq.~\eqref{sigma1}. In this case, the first-order equation \eqref{fokstatic} leads to the same solution in Eq.~\eqref{solD1}. The energy density in Eq.~\eqref{t00kstatic} is
\be\label{t00D1k}
\tensor{\Tc}{^0_0} = \frac{\sqrt{1+a}\,\sech^{4n}(x)}{\left(1+a\tanh^2(x)\right)^{(4n+1)/2}}.
\ee
In Fig.~\ref{figD1k}, we display the above energy density, with the notation $\rho=\tensor{\Tc}{^0_0}$  for some values of the parameters. Notice that the behavior is quite similar to the canonical model ($n=1$), with an internal structure arising as we have $a\to-1$, but the thickness gets smaller as $n$ increases.
The energy can be obtained by integrating the above expression in all space; it is $E=2^{4n-1}\Gamma^2(2n)/\Gamma(4n)$.

The stability potential \eqref{ukstatic} reads
\be
\begin{aligned}
    U(x)\! &=\! \frac{4(2n -1)(a+1) \tanh^2(x)}{1+a\tanh^2(x)}-\frac{2(2 n -1) \sqrt{a+1}}{\left(1+a\tanh^2(x)\right)^{3/2}}\,\\
    &\cdot \left((4a(n-1) +4 n -3) \sech^2(x)-4(n-1)(a +1)\right).
\end{aligned}
\ee
The above expression recovers the corresponding one for the standard case in Eq.~\eqref{ustdD1} for $n=1$. Since the solution and the impurity do not depend on $n$, the zero mode \eqref{zeromode} associated with the above stability potential also does not. Moreover, $\eta_0$ is the very same as in Eq.~\eqref{etazeroD1}. It does not present nodes, as we have seen in the previous section, ensuring the linear stability of the solution.

\subsection{Born-Infeld}\label{secbi}
In the previous section, we have considered that the dynamical term in the Lagrangian density is of power-law type, without coupling to functions of the scalar field. In this section, we investigate the Born-Infeld-like model described in Ref.~\cite{brax}, but now coupled to impurities, in the form
\be
F(\phi,X)= -V(\phi)\left(1-2X\right)^a,
\ee
where $a\geq1/2$. In the above expression, the field and its derivatives are coupled. From Eq.~\eqref{momentuml} for the above function, one gets $\Pi^\mu = 2aV(1-2X)^{a-1}(\partial^\mu\phi+\sigma^\mu)$. The equation of motion \eqref{eoml} associated to this model is 
\be
\partial_\mu\Pi^\mu = -V_\phi\left(1-2X\right)^a + \sigma_\mu G^\mu_\phi.
\ee
To have null divergence in the tensor \eqref{emt}, we take the constraint in Eq.~\eqref{constraint}, which reads
\be\label{constrbi1}
-2a(1-2X)^{2a-1}(V^2)_\phi = (G_\mu G^\mu)_\phi.
\ee
This constraint must be solved together with Eq.~\eqref{XG}, which we now write as
\be\label{constrbi2}
8a^2X(1-2X)^{2a-2}V^2=G_\mu G^\mu.
\ee
Solving the last system formed by the last two equations to determine $X$ and $V(\phi)$ is not easy. However, one can show that a possible solution is to take
\be\label{bicovgenv}
V^2(\phi) = -\frac{(2a-1)^{2a-1}}{(2a)^{2a}}\,G_\mu G^\mu
\ee
and
\be
X=-\frac{1}{2(2a-1)}.
\ee
These expressions allow us to get the first-order equation \eqref{fogeral} as
\be\label{bicovgenfo}
\partial_\mu\phi=-\sigma_\mu - \frac{G_\mu}{\sqrt{-(2a-1)G_\alpha G^\alpha}}.
\ee

Notice that the case $a=1/2$, which represents the so-called tachyonic dynamics \cite{tach1,tach2,tach3,tach4}, is not compatible with equations \eqref{bicovgenv} -- \eqref{bicovgenfo}. In this specfic situation, the constraints \eqref{constrbi1} and \eqref{constrbi2} are satisfied by
\be\label{bicov12v}
V(\phi) = \sqrt{-\,G_\mu G^\mu + C}\quad\text{and}\quad X = \frac{1}{2C}\,G_\mu G^\mu,
\ee
where $C$ is a constant of integration. The first-order equation \eqref{fogeral} associated with this model is
\be\label{bicov12fo}
\sqrt{C}\,(\partial_\mu\phi+\sigma_\mu) + G_\mu=0.
\ee
Notice that the integration constant is present in the latter expressions.

\subsubsection{Static case}
Let us now consider the case of static configurations. Here, the equations which describe the model also depend on $a$. For $a>1/2$, the first-order equation \eqref{bicovgenfo} simplifies to
\be\label{fostaticbi}
\nabla\phi = \vec{\sigma}+\frac{1}{\sqrt{2a-1}}\,\frac{\vec{W}_\phi}{\big|\vec{W}_\phi\big|},
\ee
where Eq.~\eqref{gw} was used. To ensure its compatibility with the equation of motion, we impose that the potential has the form \eqref{bicovgenv}, which reads
\be\label{vabi}
V(\phi) = \sqrt{\frac{(2a-1)^{2a-1}}{(2a)^{2a}}}\,\big|\vec{W}_\phi\big|.
\ee
 The energy density obtained from Eqs.~\eqref{rhonablaphi} and \eqref{rhophi} take the form
\be\label{t00abi}
    \tensor{\Tc}{^0_0} = \frac{1}{\sqrt{(2a-1)}}\,\big|\vec{W}_\phi\big| + \vec{\sigma}\cdot\vec{W}_\phi.
\ee
The stability against contractions and dilations requires the condition in Eq.~\eqref{condderrickf2}, which reads
\be\label{condderrickbi}
\begin{aligned}
 &\int_{\mathbb{R}^D} d^Dx\,\sqrt{2a-1}\,\big|\vec{W}_\phi\big|\Bigg\{ \big(\nabla\phi+(\vec{x}\cdot\nabla)\,\vec{\sigma}\big)^2\\
    &-\frac{(1-a)(2a-1)}{a}\Big[\big(\nabla\phi-\vec{\sigma}\big)\cdot\big(\nabla\phi+(\vec{x}\cdot\nabla)\,\vec{\sigma}\big)\Big]^2\Bigg\}>0,
\end{aligned}
\ee
To analyze the stability against spatial translations and small fluctuations, we must consider the matrix components in Eq.~\eqref{mijgeral}, which take the form
\be\label{mijbi}
\Mc_{ij}=\delta_{ij} + \frac{a-1}{a}\,\frac{W_\phi^i W_\phi^j}{\big|\vec{W}_\phi\big|^2}.
\ee
To ensure that the minimum energy is attained at the location of the solution, i.e., the solution is stable under spatial translations, we must take the Hessian matrix \eqref{stabtransf} into account, whose components are now given by
\be\label{stabtransbi}
\begin{aligned}
	{\cal H}_{ij} &= \int_{\mathbb{R}^D}d^Dx\,\sqrt{2a-1}\,\big|\vec{W}_\phi\big|\Bigg[\partial_i\vec{\sigma}\cdot\partial_j\vec{\sigma}\\
	&-\frac{1-a}{a}\frac{\big(\vec{W}_\phi\cdot\partial_i\vec{\sigma}\big)\,\big(\vec{W}_\phi\cdot\partial_j\vec{\sigma}\big)}{\big|\vec{W}_\phi\big|^2}\Bigg].
\end{aligned}
\ee
The solution is stable if $\text{det}({\cal H})>0$ and at least one of the components of the diagonal of ${\cal H}_{ij}$ is positive. Lastly, we look into the stability under small fluctuations, which is governed by the eigenvalue equation \eqref{stabstatic} for $\Mc_{ij}$ in Eq.~\eqref{mijbi} and the stability potential \eqref{stabpot} given by
\be\label{stabbia}
\begin{aligned}
    U(\vec{x})&=\frac{\vec{\sigma}}{\sqrt{2a-1}\,\big|\vec{W}_\phi\big|}\,\cdot\Bigg[\vec{W}_{\phi\phi\phi}-\Bigg(\frac{\big|\vec{W}_\phi\big|_\phi}{\big|\vec{W}_\phi\big|}\,\vec{W}_\phi\Bigg)_{\phi}\Bigg].
\end{aligned}
\ee
This stability potential is null in the absence of impurities, agreeing with the impurity-free case investigated in Ref.~\cite{trilogia1}. The first-order equation \eqref{fostaticbi} is quite intricate in arbitrary dimensions; its last term presents sign function which depends on the solution itself. This investigation will not be done here.

\subsubsection{Single spatial dimension}
\begin{figure}[t!]
    \centering
\includegraphics[width=0.6\linewidth]{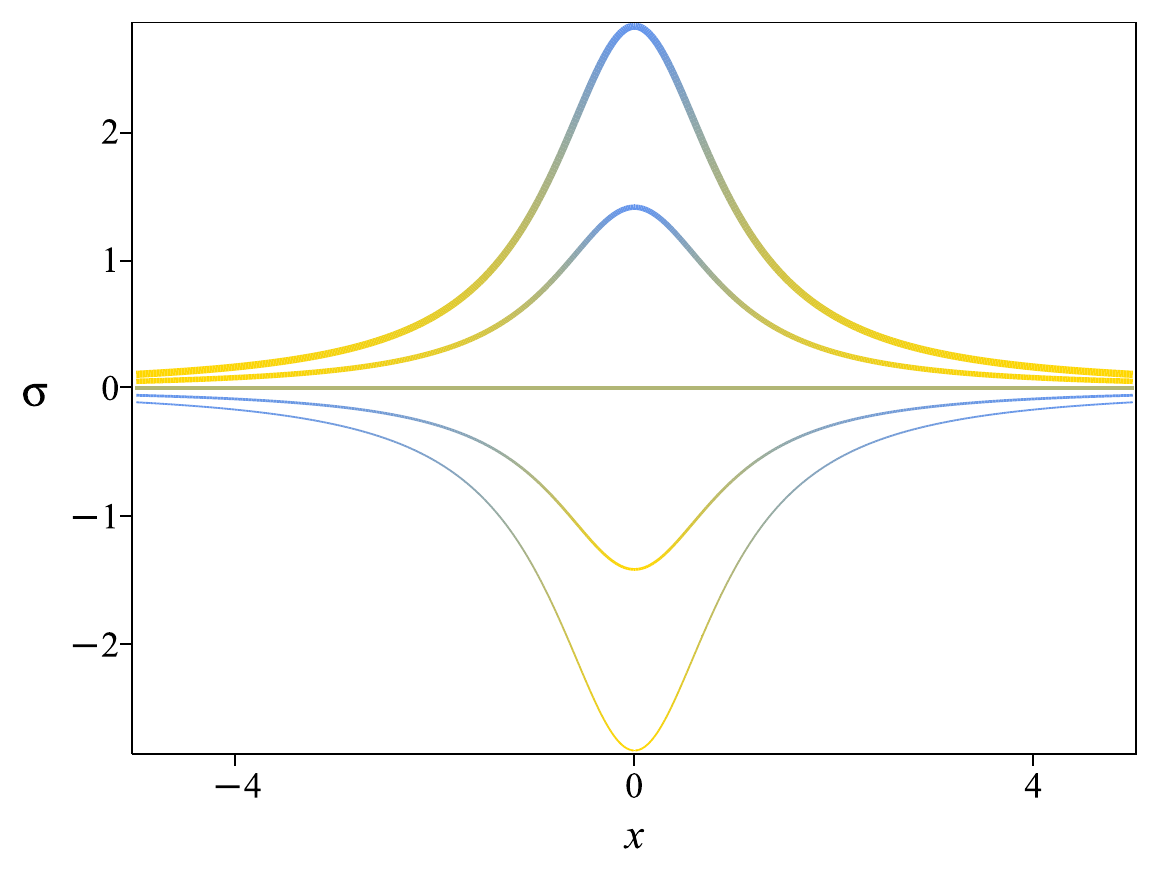}
\includegraphics[width=0.6\linewidth]{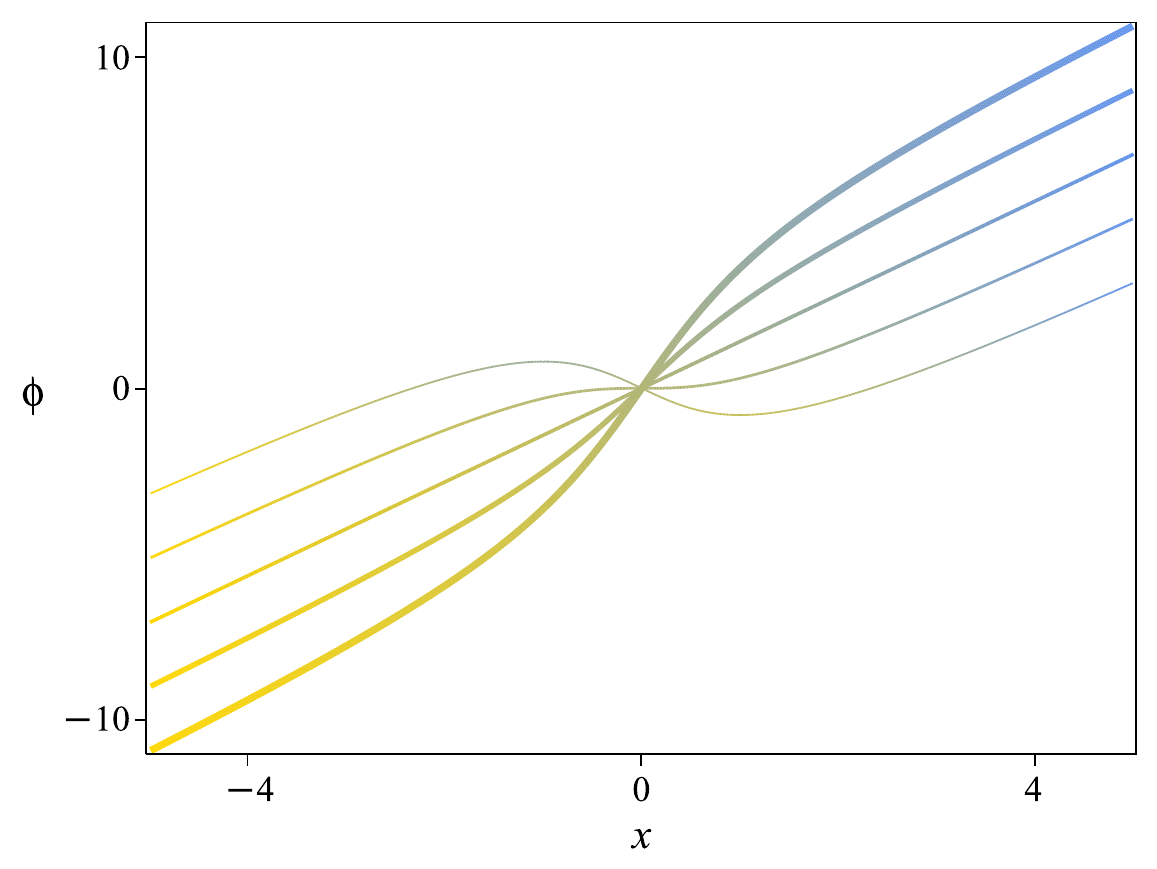}
\includegraphics[width=0.6\linewidth]{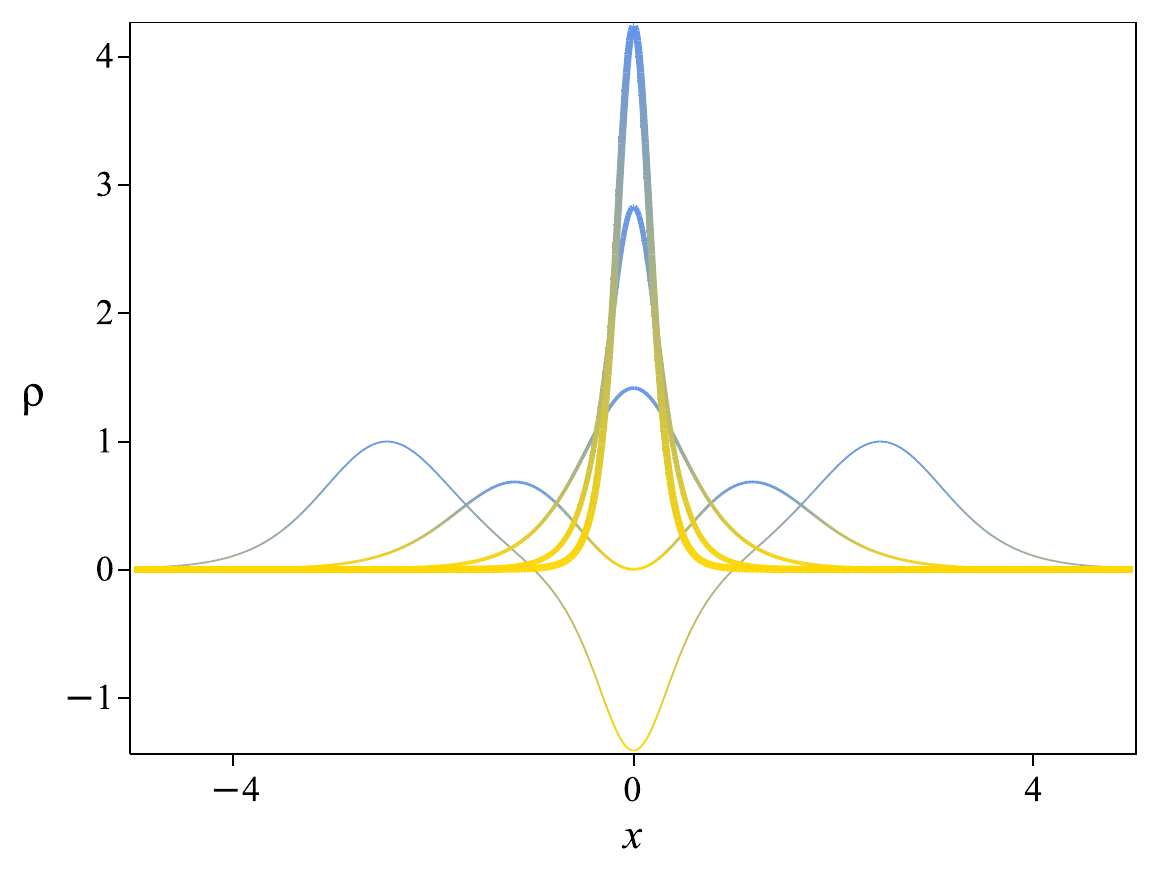}
    \caption{The impurity in Eq.~\eqref{modelbi} (top), the solution \eqref{solbiaD1} (middle) and the energy density $\rho=\tensor{\Tc}{^0_0}$ in Eq.~\eqref{rhobiaD1} (bottom) for $a=3/4$ and $\alpha=-2\sqrt{2},-\sqrt{2},0,\sqrt{2}$ and $2\sqrt{2}$. The thickness of the lines increases with $\alpha$.}
    \label{figbi}
\end{figure}

To illustrate our procedure, we only deal with the case $D=1$ explicitly, where this issue can be avoided by considering the case in which $W_\phi$ is non negative, so the first-order equation \eqref{fostaticbi} admits the solution
\be\label{solbigeralD1}
\phi(x) = \frac{x}{\sqrt{2a-1}} + \int dx\,\sigma(x),
\ee
in which a constant of integration always arises. It is worth commenting that this solution can be used in Eq.~\eqref{zeromode} to get that the zero mode of the stability equation is always constant. In the absence of impurities ($\sigma=0$), the above expression also shows us that the solution is always a straight line. The energy density is
\be
    \tensor{\Tc}{^0_0} = \frac{|W_\phi|}{\sqrt{2a-1}} + \sigma(x) W_\phi.
\ee
We then see that the solution and the energy density depend on the auxiliary function and the impurity. To illustrate the model, we take
\be\label{modelbi}
W(\phi) = \tanh(\phi)\quad\text{and}\quad\sigma(x)=\frac{\alpha}{1+x^2},
\ee
so the potential \eqref{vabi} has the form 
\be
V(\phi) = \sqrt{\frac{(2a-1)^{2a-1}}{(2a)^{2a}}}\,\sech^2(\phi).
\ee
By using the functions in Eq.~\eqref{modelbi} into the general solution \eqref{solbigeralD1}, we get
\be\label{solbiaD1}
\phi(x) = \frac{x}{\sqrt{2a-1}} + \alpha\arctan(x).
\ee
The case $\alpha=0$ recovers the impurity-free ($\sigma=0$) model. Near the origin, for $x\approx0$ one can show that it has the form $\phi(x)\approx \left(\alpha + 1/\sqrt{1-2a}\right) x - \alpha x^3/3 + {\cal O}[x^5]$. Therefore, the impurity modifies the slope of the solution at the origin via the parameter $\alpha$. Asymptotically, for $x\to\pm\infty$, the solution behaves as $\phi_\pm(x) \approx x/\sqrt{2a-1} \pm \alpha\pi/2 \mp\alpha/x + {\cal O}[1/x^3]$. The energy density associated to the above solution is
\be\label{rhobiaD1}
\begin{aligned}
    \tensor{\Tc}{^0_0}(x) &= \left(\frac{1}{\sqrt{2a-1}} +\frac{\alpha}{1+x^2}\right)\\
    &\cdot\sech^2\left(\frac{x}{\sqrt{2a-1}} + \alpha\arctan(x)\right).
\end{aligned}
\ee
By integrating the above expression, we get that the energy is $E=2$, as expected from Eq.~\eqref{ewD1}, being independent of $a$ and $\alpha$. For $\alpha>\alpha_*=-1/\sqrt{2a-1}$, the solution is monotonically increasing and the energy density has a bell shape. If $\alpha=\alpha_*$, the solution gets null slope and the energy density presents a hole at $x=0$ with two peaks around this point. The case $\alpha<\alpha_*$ presents an exotic feature: the slope of the solution is negative; this makes the hole in the energy density become deeper, and the peaks go away from the origin. In Fig.~\ref{figbi}, we display the impurity in Eq.~\eqref{modelbi}, the solution \eqref{solbiaD1}, and the above energy density for some values of the parameters.

The stability of the solution \eqref{solbiaD1} against rescaling of argument is given by Eq.~\eqref{condderrickbi}. By using numerical methods, we have shown that it is satisfied, so the solution is stable against contractions and dilations. The stability against translations in the space is related to Eq.~\eqref{stabtransbi}. Since we are working in a single spatial dimension, the only surviving component of the Hessian matrix is ${\cal H}_{11} = \int_{-\infty}^{\infty} dx\,((2a-1)^{3/2}/a)\,W_\phi{\sigma^\prime}^2$, which is positive for the auxiliary function and the impurity in Eq.~\eqref{modelbi}. This ensures that the solution \eqref{solbiaD1} will not translate in the space spontaneously.

The stability potential in Eq.~\eqref{stabbia} associated to the stability under small fluctuations gets the form
\be
\begin{aligned}
    U(x) &= \frac{6\alpha}{\sqrt{2a-1}}\,\frac{1}{\left(1+x^2\right)}\\
    &\cdot\tanh^2\left(\frac{x}{\sqrt{2a-1}} + \alpha\arctan(x)\right).
\end{aligned}
\ee
The zero mode is given by Eq.~\eqref{zeromode}. For the solution \eqref{solbiaD1}, one can show that even though it exists, it is uniform in all the space, so one cannot normalize it.

We remark that the solution in Eq.~\eqref{solbiaD1} was obtained for the spatially localized impurity in Eq.~\eqref{modelbi}. However, the first-order equation for general $W_\phi$ and $\sigma$ opens other possibilities up, in which the impurity does not vanish asymptotically and $W_\phi$ calculated in the solution may change the sign. This can lead to solutions with other features, such as distinct asymptotic behavior. 

For $a=1/2$, we obtain from Eqs.~\eqref{bicov12fo} and \eqref{bicov12v} the following first-order equation and potential, respectively,
\be
\nabla\phi=\vec{\sigma} + \frac1{\sqrt{C}}\,\vec{W}_\phi,\quad V(\phi) =\sqrt{C+\big|\vec{W}_\phi\big|^2}.
\ee
The energy density can be calculated from Eq.~\eqref{t00static}, which reads
\be
\tensor{\Tc}{^0_0}=V(\phi)\sqrt{1-2X} + \vec{\sigma}\cdot\vec{W}_\phi,
\ee
By using the above first-order equation and potential, one can show that $\tensor{\Tc}{^0_0}=\nabla\cdot\vec{W}+\sqrt{C}$. Therefore, to ensure that the energy is finite, we impose $C=0$. Interestingly, one can show by similar arguments that the absence of $C$ is also required by Eq.~\eqref{derrick1a} to avoid instabilities against contractions and dilations. We highlight that we now have $V(\phi)=|\vec{W}_\phi|$ and we must analyze Eq.~\eqref{constrbi2} as the first-order equation is not valid for $C=0$. We then get
\be
\frac{|\vec{W}_\phi|}{1+\left|\nabla\phi-\vec{\sigma}\right|^2}=0.
\ee
The only way to satisfy this equation is by taking $|\nabla\phi-\vec{\sigma}|\to\infty$ in the region where $\vec{W}_\phi\neq0$ and $|\nabla\phi-\vec{\sigma}|=$ constant where $\vec{W}_\phi$ vanishes. For $D=1$ and $\sigma=0$, this gives rise to the so-called tachyon kink \cite{tach1,tach2,tach3,tach4}. The study of linear stability of singular tachyon kinks is already intricate in the absence of impurities. We do not investigate this issue here, as it may require a special approach that differs from the one that we have used.

\section{Final Remarks}\label{secfinal}
In this work, we have investigated the conditions that must be imposed in Lagrangian densities associated to a single real scalar field with explicit dependence on the spacetime coordinates to get the tensor \eqref{emt} with null divergence. We have introduced the class of models \eqref{lagrangian} that engenders impurities but still allows for $\partial_\mu\tensor{\Tc}{^\mu_\nu}=0$. Interestingly, the formalism have led us to a first-order equation which is compatible with the equation of motion if a constraint is satisfied.

Our formalism can be applied to time-dependent fields and impurities. However, to better understand how it works, we considered the case with static field and impurity. In this situation, we have investigated the stability against contractions and dilations, which is associated to the pressures and requires the conditions \eqref{derrick1}, and also, the stability against spatial translations, which is related to the Hessian matrix \eqref{stabtrans}. These stabilities can only be attained if the condition which ensures the divergenceless character of the tensor $\tensor{\Tc}{^\mu_\nu}$ is satisfied.

In the specific class of models \eqref{lagrangian}, our formalism has led to introduce the auxiliary vector function $\vec{W}(\phi)$ that allows us to write the energy as the integral of a divergence. By considering that $\vec{G}=\vec{W}_\phi$, we have shown that stable solutions require null energy for $D>1$. The case $D=1$ is special and the value of the energy depends on the specific model under investigation. We have also analyzed the stability of the static solutions against small fluctuations in arbitrary dimensions. It is governed by a partial differential eigenvalue equation whose associated operator can only be factorized in terms of adjoint operators if the static field solves the first-order equation that emerges from our formalism. We have shown that the zero mode may not exist, as it is described by a partial differential equation that may not support solutions.

The static case in the model \eqref{lagrangian} was also investigated in a single spatial dimension. Interestingly, Derrick's scaling argument lead to a \emph{local} condition, contrary to the case in higher dimensions. Moreover, the conditions required to preserve the hyperbolicity of the eigenvalue equation that governs the stability under small fluctuations makes the requirements to get stability against translations or contractions and dilations become identities. Even so, the linear stability is not ensured; it depends on the model of interest. We have used the first-order equation to factorize the Sturm-Liouvile operator associated to the eigenvalue equation in terms of adjoint operators to show that the zero mode can be obtained analytically. The solution is stable only if the zero mode does not present nodes.

To show that our procedure is robust, we have first developed it for the canonical model coupled with impurities. We have constructed the potential that allows for the presence of the first-order equation which ensures the equality $\partial_\mu\tensor{\Tc}{^\mu_\nu}=0$ for \emph{time-dependent} field and impurity in arbitrary dimensions. In the static case, the conditions that ensure stability against rescaling become identities. It is worth to highlight that the model supports a BPS bound if the static versions of the aforementioned potential and first-order equations are used. In this situation, the operator that dictates the linear stability can be factorized explicitly in terms of adjoint operators. If these operators do not present divergences, then the negative eigenvalues are absent and the model is linearly stable. We illustrate the method for the specific set of potential and impurity considered in Ref.~\cite{letterimpurity} in two spatial dimensions, in which an analytical solution is possible. It connects the minima of the potential and engenders null energy, as required by stability conditions. Also, we have shown numerically that the condition for avoiding instabilities against translations is satisfied. In a single spatial dimension, we have also considered the solution studied in Ref.~\cite{letterimpurity}. The energy is finite and does not depend on the aforementioned parameter. We have studied the linear stability, obtaining the stability potential and calculating the zero mode explicitly; it does not have nodes, confirming the stability of the solution.

Since our formalism allows us to study non-canonical models, we have considered a model in which the potential is not coupled to the derivatives of the field in the Lagrangian density, but the dynamical term contains a generalization, known as $k$-fields, similar to \cite{babichev,adamk,trilogia2}, described by a term of a power-law type of exponent $n$. In this case, we have shown the explicit form of the potential that satisfies the constraint required to compatibilize the equation of motion with the first-order equation that leads to the null divergence of $\tensor{\Tc}{^\mu_\nu}$ in the \emph{time-dependent} $(D,1)$-dimensional scenario. To study the properties of the model, we then have considered the static case, presenting the general equations with the auxiliary function defined in Eq.~\eqref{gw}. In two and one spatial dimensions, by considering specific forms for the auxiliary function $\vec{W}(\phi)$, we have shown that the same solutions of the canonical model can be obtained, provided that the impurity is also the same. The energy density, although has a similar profile, depends on the parameter $n$, which changes the thickness of the solutions. We have shown that the stabilities against spatial translation and rescale of argument are ensured. The $D=2$ case has an issue: the stability potential ranges from $-\infty$ to $\infty$, with a continuum of negative modes being present. This leads to instability. In a single spatial dimension, we have obtained the potential that describes the linear stability and the zero mode, which is the very same of the canonical model; it does not engender modes, ensuring the stability under small fluctuations.

Another generalized model which we have considered was of the Born-Infeld type. It couples the field to its derivative in a non-trivial manner, with the presence of the parameter $a$. We have found a class of models that allows for $\partial_\mu\tensor{\Tc}{^\mu_\nu}=0$. In the static case, we have investigated the aforementioned instabilities and showed that the first-order equations become quite intricate when compared to the impurity-free scenario. However, we have investigated the model with $a>1/2$ in a single spatial dimension and obtained analytical solutions for a spatially localized impurity. The results show that the impurity modifies the slope of the solution, which may break the monotonic profile, leading to an interesting internal structure to the solution and energy density. In this situation, the zero mode is uniform in all the space.

There are several distinct possibilities of continuation of the present study. In particular, we can consider the case of time-dependent field and impurities and also, the use of several fields. Further investigation of the linear stability of impurity-doped models in arbitrary dimensions are also desirable. For instance, one may consider other $\vec{W}$ and $\vec{\sigma}$ in the context of $k$-field models to seek linearly stable solutions in $D=2$ and $D=3$. It is also of interest to describe time-dependent complex scalar field, to examine if the above methodology can be extended to Q-balls. We may also think of using complex scalar fields coupled to a gauge field engendering local $U(1)$ symmetry to investigate vortices in the plane along the lines of Refs. \cite{vimp1,vimp2,vimp3,vimp4,vimp5,vimp6,vimp7} and also, models with $SU(2)$ symmetry coupling triplets of scalar and gauge fields to describe magnetic monopoles in space. Moreover, if one considers two scalar fields, one complex and the other real, it is possible to examine generalization of the Friedberg-Lee-Sirlin model \cite{FLS} to incorporate impurities in three spatial dimensions. Another line of investigation concerns studying fields and impurities on curved backgrounds \cite{Morris} and also, extensions including the Einstein-Friedberg-Lee-Sirlin model, which has been recently considered to investigate black holes, Q-balls, and boson stars \cite{blackhole,Qballs,bosonstars}.

\acknowledgements{This work is supported by the Brazilian agency Conselho Nacional de Desenvolvimento Cient\'ifico e Tecnol\'ogico (CNPq), grants Nos. 402830/2023-7 (DB, MAM and RM),  303469/2019-6 (DB), 306151/2022-7 (MAM) and 310994/2021-7 (RM).}

\end{document}